\documentclass[11pt,a4paper]{article}
\pdfoutput=1

\usepackage[a4paper,text={16.8cm,22.4cm}]{geometry}
\usepackage{amsmath,amsfonts,slashed,amssymb,tikz,bm,psfrag,graphicx,color,dsfont}
\usepackage{multicol}
\usepackage{float}
\usepackage{slashed}
\usepackage{euscript}
\usepackage[normalem]{ulem}
\RequirePackage[sort&compress,square,comma,numbers]{natbib}
\allowdisplaybreaks
\renewcommand{\arraystretch}{1.2}
\usepackage{hyperref}
\usepackage{graphicx}
\usepackage{amsmath}
\usepackage{amsfonts}      % facilitate mathematical typesetting
\usepackage{empheq}       % box equations
\usepackage{hyperref}     % hyperlinks
\usepackage{euscript}
\usepackage{multirow}
\usepackage{simplewick}
\usepackage{xcolor}
\usepackage{slashed}

\newcommand{\decay}{B \to \gamma \ell \nu_{\ell}}
\def\slash#1{#1 \hskip-0.53em /}

\allowdisplaybreaks

\definecolor{brown}{RGB}{150,50,0}
\begin{document}

\begin{titlepage}

\begin{flushright}
\normalsize

TUM-HEP-1330/21
\end{flushright}

\vspace{0.1cm}
\begin{center}
\Large\bf
QCD factorization for the $\decay$ decay beyond leading power
%$B\to \gamma\ell\nu$ decay in QCD
\end{center}

\vspace{0.5cm}
\begin{center}

{\bf Bo-Yan Cui$^{a,b}$\,, Yue-Long Shen$^{c}$\,,
Chao Wang$^{d}$ and  Yan-Bing Wei$^{\ast,e,f,g}$} \\
\vspace{0.5cm}
{\sl
${}^a$ Department of Physics, Chongqing University of Science and Technology,\\
Chongqing 401331, P.R. China

${}^b$ School of Physics, Nankai University, Tianjin 300071, P.R. China \\
${}^c$ College of Information Science and Engineering,
Ocean University of China, Songling Road 238, Qingdao, Shandong 266100, P.R. China\\
${}^d$ Faculty of Mathematics and Physics, Huaiyin Institute of Technology,
Huaian, Jiangsu 223001, P.R. China\\
${}^e$ Institute of Particle Physics and Key Laboratory of Quark and Lepton Physics (MOE),
Central China Normal University, Wuhan, Hubei 430079, China\\
${}^f$ Faculty of Science,  Beijing University of Technology,
Beijing 100124, P.R. China\\
${}^g$Physik Department T31, James-Franck-Stra{\ss}e 1, Technische Universit\"at M\"unchen, \\
D-85748 Garching, Germany
}\\
\vspace{0.2cm}
{\sl E}-{\it mail}: {\tt boyancui@cqust.edu.cn, shenylmeteor@ouc.edu.cn, \\
chaowang@nankai.edu.cn,  yanbing.wei@ccnu.edu.cn}
\vspace{0.5cm}
%\today
\end{center}

%%%%%%%%%%%%%%%%%%%%%%%%%%%%%%%%%

%\subheader{
%\begin{flushright}
%{\small
%TUM-HEP-1330/21\\
%arXiv:1804.04962 [hep-ph]\\
%\today
%}
%\end{flushright}
%}

\begin{abstract}
The radiative leptonic $\decay$ decay  serves as an ideal platform to determine the $B$-meson inverse moment which is a fundamental nonperturbative parameter for the $B$ meson.
In this paper, we explore precise QCD contributions to this decay with an energetic photon.
We reproduce the next-to-next-to-leading-logarithmic resummation formula for the decay amplitude at leading power in $\Lambda_{\rm QCD}/m_b$.
Employing operator identities,
we calculate subleading-power contributions from the expansion of the hard-collinear propagator of the internal up quark and the heavy-quark expansion of the bottom quark.
We update the contributions from the hadronic structure of the photon to the $\decay$ process with the dispersion technique.
Together with other yet known power corrections, phenomenological applications including the partial branching fraction and ratio of the branching fractions of the radiative $B$ decay are investigated.
\end{abstract}
\vfil

\begin{flushleft}
\normalsize

$^\ast$Corresponding author.
\end{flushleft}
\end{titlepage}
%\keywords{B physics, NLO computations, nonperturbative effects}

%\arxivnumber{1234.5678}https://www.overleaf.com/project/60bb8172db0a9d24a9bf3e3f

\tableofcontents

%%%% not showing subsubsections in the table of contents
\setcounter{tocdepth}{2}
%%%%

%\notoc

\makeatletter
\gdef\@fpheader{}
\makeatother

\flushbottom

\section{Introduction}\label{sec:introduction}
The radiative leptonic $\decay$ decay is of great interest
for its offering a clean probe for determining $\lambda_B$, the inverse moment of $B$-meson light-cone distribution amplitude (LCDA), which is a crucial
nonperturbative parameter for the $B$ meson in the heavy-quark effective theory (HQET).
Employing the Standard-Model prediction for the radiative $B$ decay~\cite{Beneke:2018wjp}, the Belle collaboration has reported a lower limit for the inverse moment $\lambda_B>0.24$~GeV at 90\% CL~\cite{Gelb:2018end}.
To extract a more practical value for the inverse moment, more accurate experimental data and theoretical calculations are indispensable.
With the ongoing super $B$ factory Belle-II collaboration that will provide greatly improved data shortly, it is urgent to re-investigate the radiative leptonic $B$ decay from the theoretical point of view.

Factorization properties of the $\decay$ decay have been
explored at leading power (LP) in $\Lambda_{\rm QCD}/m_b$ with distinct
QCD techniques including the soft-collinear effective theory (SCET)~\cite{Korchemsky:1999qb,DescotesGenon:2002mw,Lunghi:2002ju,Bosch:2003fc,Beneke:2003pa}, and the LP contribution to the decay with the short-distance hard-scattering kernel up to the next-to-next-to-leading-order (NNLO) accuracy in $\alpha_s$ was considered recently~\cite{Galda:2020epp,Galda:2022dhp}.
Also next-to-leading-power (NLP) corrections in $\Lambda_{\rm QCD}/m_b$ to the $\decay$ decay were discussed extensively~\cite{Beneke:2011nf,Braun:2012kp,Wang:2016qii,Wang:2018wfj,Beneke:2018wjp,Shen:2020hsp,Janowski:2021yvz}.
The NLP contributions include the local contribution from subleading terms of the hard-collinear quark propagator and the correction from photon emitted off the $b$ quark both at leading order (LO) in $\alpha_s$~\cite{Beneke:2011nf}, the soft contribution with the light-cone sum rules (LCSR) approach up to next-to-leading order (NLO) in $\alpha_s$~\cite{Braun:2012kp,Wang:2016qii}, the tree-level contribution with higher-twist $B$-meson LCDAs~\cite{Wang:2016qii,Beneke:2018wjp}, the LO correction from subleading terms of the heavy-quark expansion of the $b$ quark~\cite{Beneke:2018wjp}, and the resolved photon contribution up to NLO~\cite{Wang:2018wfj}.

Motivated by the recently derived LP factorization formula for the $\decay$ decay and
various NLP corrections to the decay, we attempt to improve the
theoretical predictions for the leptonic radiative $B$ decay. The main new ingredients of the present work
are summarized as follows:
\begin{itemize}
%\item
%\sout{Applying the renormalization-group (RG) equation approach, we obtain formally the LP factorization formula for the $B\to \gamma\ell\nu$ decay up to the next-to-next-to-leading-logarithmic (NNLL) accuracy.
%Once the yet unknown three-loop anomalous dimension of the $B$-meson LCDA is calculated, the factorization formula can be employed to resum large logarithms in it to the NNLL level directly.}
\item
We calculate the ``non-local'' NLP correction from the hard-collinear quark propagator at the tree level by using operator-level identities.
The operator identities are indispensable for non-local operators whose fields do not have light-like separation.
Further, employing operator-level identities, we re-derive the LO contribution from subleading terms in the heavy-quark expansion of the $b$ quark field.

\item
We update the resolved photon corrections in the framework of LCSR with photon LCDAs.
Employing the proper momentum power-counting scheme, we calculate the leading-twist photon-LCDA contribution to the NLO accuracy.
Also, we calculate the LO corrections from higher-twist photon LCDAs with the correct definition of these LCDAs.

\end{itemize}
We find that the newly calculated resolved photon contribution and the correction from the hard-collinear quark propagator are also numerically important (see Section~\ref{sec:Numrical}).

The rest of this paper is organized as follows:
We first give a prefatory overview of the
$\decay$ process in Section~\ref{sec:theory}.
In the following section, we collect the NNLL-resummation formula for the decay amplitude at LP.
Then we calculate various subleading-power corrections in Section~\ref{sec:NLP}.
Further in Section~\ref{sec:Numrical}, numerical impacts of higher-order contributions  in $\alpha_{s}$ and
higher-power contributions in $\Lambda_{\rm QCD}/m_{b}$ are discussed.
Finally, a summary of the main conclusion of this work is given in
Section~\ref{sec:Conclusion}.

\section{Theoretical overview of the $\decay$ decay}
\label{sec:theory}

The $\decay$ decay amplitude can be defined by the QCD matrix element
\begin{align}\label{DecayAmplitude}
\mathcal{A}(B^-\to\gamma \ell \bar{\nu})=
\frac{G_F\,V_{ub}}{\sqrt{2}}
\langle \gamma\,\ell\,\bar{\nu} |\,
[\bar{\ell}\gamma_\mu(1-\gamma_5)\nu]\,
[\bar{u}\gamma^\mu(1-\gamma_5)b]\,
| B^- \rangle\;.
\end{align}
We will work in the $B$-meson rest frame in which the momentum of the $B$ meson is $m_B v$ with the four vector $v_\mu=(1,0,0,0)$.
With the photon having momentum $p$, the momentum transfer between the $B$ meson and the photon $q=m_B v-p$ represents the sum of momenta of the final state leptons.
Since there exist light-like final state particles, it is convenient to employ the light-cone coordinate by introducing two light-cone vectors $n_{\mu}$ and $\bar{n}_{\mu}$ satisfying $n^{2}=\bar{n}^{2}=0$ and $n\cdot \bar{n}=2$.
In this coordinate we have $v_\mu =(n_{\mu}+\bar{n}_{\mu})/2$,
and a given vector could be decomposed as
$k_{\mu}=n\cdot k\,\bar{n}_{\mu}/2
+\bar{n}\cdot k\,n_{\mu}/2 + k_{\perp\mu}$.
Setting the photon propagating along the $\bar{n}$ direction, the relevant momenta are expressed as
\begin{align}\label{MomentumDef}
p_\mu
=&~\frac{n\cdot p}{2}\,\bar{n}_\mu
\equiv E_\gamma \bar{n}_\mu\,,
&
q_\mu=&~\frac{n\cdot q}{2}\, \bar{n}_\mu
+\frac{\bar{n}\cdot q}{2} \, n_\mu\,,
\end{align}
with $E_{\gamma}\le m_B/2$ the energy of the photon.

Taking advantage of the Ward identity, the radiative $B$ decay is determined by two form factors, $F_{V}$ and $\hat{F}_{A}$, defined through the vacuum-to-$B$-meson correlation function
\begin{align}
\int d^{4}x \, e^{ip\cdot x} \,
\langle 0  | \,
{\rm T} \{ \, j^{\nu}_{\rm em}(x),
[\bar{u}\gamma^\mu(1-\gamma_5)b](0)
\}
\,|  B^{-} \rangle
=  v\cdot p \, \big( \,
g^{\mu\nu} \, \hat{F}_{A}
-i \,\epsilon^{\mu\nu nv}\, F_{V}\big) +\cdots \,,
\label{eq:correlation.function}
\end{align}
where the electromagnetic current $j^{\nu}_{\rm em}=\sum_q Q_q \, \bar{q} \gamma^\nu q$
%+\sum_\ell Q_\ell \, \bar{\ell} \gamma^\nu \ell$
and $\epsilon^{0123}=1$.
Note that the form factors for the other Lorentz structures, which are not crucial for the radiative leptonic $B$ decay, are represented by the ellipsis (see Ref.~\cite{Wang:2016qii} for details).
The differential decay width of the $\decay$ decay is
\begin{align}
\frac{d \, \Gamma}{d \, E_{\gamma}}
=&~ \frac{\alpha_{\rm em} \, G_F^2 \, |V_{ub}|^2}{6\, \pi^2} \,
m_B \, E_{\gamma}^3 \,
\left (  1- \frac{ 2 \, E_{\gamma}}{m_B} \right ) \,
\Big(\big|F_V\big|^2
+ \big|F_A\big|^2   \Big) \,,
\nonumber \\
F_A(E_{\gamma})=&~
\hat{F}_A(E_{\gamma}) + \frac{Q_\ell\,f_B}{E_{\gamma}}\,,
\label{eq:br}
\end{align}
where the $Q_\ell$ term in the second equation corresponds to the contribution from the photon emitted off the charged lepton.
We have neglected masses of the final state leptons.

For an energetic photon with $E_{\gamma}\sim \mathcal{O}(m_{b})$, the form factors are expressed as~\cite{Beneke:2011nf}
\begin{align}
F_{V}(E_{\gamma}) = &~ \frac{Q_{u}\,m_{B}}{2E_{\gamma}} \,
R(E_{\gamma})
+ \Big[ \xi(E_{\gamma}) + \Delta \xi(E_{\gamma}) \Big]\,,
\nonumber \\
\hat{F}_{A}(E_{\gamma}) = &~
\frac{Q_{u}\,m_{B}}{2E_{\gamma}} \,
R(E_{\gamma})
+ \Big[ \xi(E_{\gamma}) - \Delta \xi(E_{\gamma}) \Big]\,.
\label{eq:form.factor}
\end{align}
The first terms on the right-hand side of the equations stand for the LP contribution which preserves the large recoil symmetry relation.
And the power corrections are included in the terms in square brackets, where we have separated the symmetry-breaking term $\Delta \xi(E_{\gamma})$ from the symmetry-preserving term $\xi(E_{\gamma})$.

\section{Form factors at LP}
\label{sec:lp}

The form factors are determined by a single function $R(E_\gamma)$ at LP in the heavy-quark expansion.
This function is factorized into convolution of short-distance hard-scattering kernel $T$ and long-distance $B$-meson LCDA $\phi_{+}$
\begin{equation}
R(E_\gamma) = \tilde{f}_B(\mu)
\int_0^\infty \frac{d\omega}{\omega}\,
T(E_\gamma,\omega,\mu) \, \phi_+(\omega,\mu) \,,
\end{equation}
where the leading-twist LCDA $\phi_+$ is defined in Appendix~\ref{DAofB} and $\omega$ corresponds to the $n$-direction momentum component of the light quark in the $B$ meson.

Since the hard-scattering kernel $T$ contains contributions of two well separated perturbative scales, which are the hard scale of $\mathcal{O}(m_b)$ and the hard-collinear scale of $\mathcal{O}(\sqrt{m_b\Lambda_{\rm QCD}})$, it could be further factorized
\begin{equation}
T(E_\gamma,\omega,\mu)=
C(E_\gamma,\mu) \, J(E_\gamma\omega,\mu)\,.
\end{equation}
Here the hard function $C$ and the jet function $J$ receiving respectively contributions from the hard-scale and the hard-collinear-scale dynamics can be calculated by matching from QCD onto SCET$_{\rm I}$ and from SCET$_{\rm I}$ onto SCET$_{\rm II}$ individually~\cite{Lunghi:2002ju,Bosch:2003fc}.
The static $B$-meson decay constant $\tilde{f}_B$, which arises form the matrix element of HQET operator, is related to the QCD $B$-meson decay constant $f_B$ by
\begin{align}
\tilde{f}_B(\mu) = K^{-1}(\mu) \, f_B \,.
\end{align}
The factor $K$ is the short-distance coefficient for the matching of the corresponding QCD weak current onto the HQET operator.
The functions $C(E_\gamma,\mu)$, $J(E_\gamma\omega,\mu)$ and $K(\mu)$ can be calculated perturbatively, and the coefficients up to NNLO accuracy are calculated in~\cite{Bonciani:2008wf, Asatrian:2008uk,Beneke:2008ei,Bell:2008ws,Liu:2020ydl,Broadhurst:1994se} and are collected in the Appendix~\ref{App:AD} for convenience.

\subsection{Resummation improved formula}

To improve the convergence behavior of the perturbative expansion of coefficient functions, large-logarithmic resummation must be performed for these functions.
Choosing the factorization scale to be of the order of the hard-collinear scale $\sqrt{\Lambda_{\rm QCD}m_b}$, the jet function $J(E_\gamma\omega,\mu)$ is free of large logarithms.
Employing the standard RG-equation approach, we fulfill the large-logarithmic resummation to functions $C(E_\gamma,\mu)$ and $\tilde{f}_B(\mu)$ to the NNLL accuracy and the LCDA $\phi_+(\omega,\mu)$ to a partial NNLL accuracy.

It is not easy to perform the large-logarithmic resummation for the $B$-meson LCDA in the momentum space even to the leading-logarithmic (LL) accuracy, because the RG equation of the momentum-space $\phi_+(\omega,\mu)$ is an integro-differential equation.
To resum the large logarithms, we will perform a Laplace transformation to the LCDA~\cite{Lange:2003ff,Galda:2020epp}
\begin{eqnarray}\label{BmesonDA:Laplace}
\tilde{\phi}_+(\eta,\widehat{\omega},\mu)=
\int_0^\infty\frac{d\omega}{\omega}\,
\Big(\frac{\omega}{\widehat{\omega}}\Big)^{-\eta}\,
\phi_+(\omega,\mu)\,.
\end{eqnarray}
In the Laplace space, the convolution of the jet function with the LCDA is relatively trivial
\begin{align}
\int_0^\infty \frac{d\omega}{\omega}\,
J(E_\gamma\omega,\mu) \, \phi_+(\omega,\mu)
= {\EuScript J}(\partial_\eta,\mu) \,
\bigg( \frac{2E_\gamma\,\widehat\omega}{\mu^2} \bigg)^\eta\,
\tilde{\phi}_+(-\eta,\widehat\omega,\mu) \Big|_{\eta \to 0}\,,
\end{align}
where ${\EuScript J}(L_p,\mu)\equiv J(E_\gamma\omega,\mu)$ with $L_p=\ln(2E_\gamma\,\omega/\mu^2)$.
Although the LCDA $\tilde{\phi}_+$ depends on the reference parameter $\widehat\omega$, the convolution is independent of this parameter.
We can in principle set $\widehat\omega$ to be any value of $\mathcal{O}(\Lambda_{\rm QCD})$.
The resummation improved function $R(E_\gamma)$ is expressed as
\begin{equation}
R(E_\gamma) =
U_1(E_\gamma,\mu_{h1},\mu) \, C(E_\gamma,\mu_{h1})\,
U_2(\mu_{h2},\mu) \, \tilde{f}_B(\mu_{h2})
\Big[ {\EuScript J}(\partial_\eta,\mu) \,
\tilde{\phi}_{+,U}(\eta,\mu_s,\mu)
\Big]\Big|_{\eta=0} \,.
\label{eq:RG.r}
\end{equation}
%{\color{red} 
The factors $U_1$ (see \ref{eq:RG.u1}) and $U_2$ (see \ref{eq:RG.u2}) are separately the evolution factors of $C(E_\gamma,\mu)$ and $\tilde f_B(\mu)$, unlike in~[9] we distinguish the two hards scales here. 
%}
The scales $\mu_{h1}$ and $\mu_{h2}$ are both hard scales of $\mathcal{O}(m_b)$ and $\mu_s$ is a soft scale of $\mathcal{O}(\Lambda_{\rm QCD})$.
These scales ensure that functions $C(E_\gamma,\mu_{h1})$, $\tilde{f}_B(\mu_{h2})$ and $\phi_+(\omega,\mu_s)$ are free of large logarithms.
The factors $U_1$ (see \eqref{eq:RG.u1}) and $U_2$ (see \eqref{eq:RG.u2}) are separately the evolution factors of $C(E_\gamma,\mu)$ and $\tilde{f}_B(\mu)$. The function $\tilde{\phi}_{+,U}$ (see \eqref{eq:RG.phi}) corresponds to the partial NNLL resummed $B$-meson LCDA in the Laplace space, which will be defined later.

\subsubsection{NNLL resummation for $C$ and $\tilde{f}_B$}

The evolution factors $U_1(E_\gamma,\mu_{h1},\mu)$ and $U_2(\mu_{h2},\mu)$ satisfy the same RG equations as $C(E_\gamma,\mu)$ and $\tilde{f}_B(\mu)$, respectively,
\begin{align}
\frac{d}{d\ln\mu}\,U_1(E_\gamma,\mu_{h1},\mu)
=&~ \Big[\Gamma_{\rm cusp}(\alpha_s)\,\ln\frac{2\,E_\gamma}{\mu}
+\gamma_H(\alpha_s)\Big]\,U_1(E_\gamma,\mu_{h1},\mu)\,,
\nonumber\\
\frac{d}{d\ln\mu}\,U_2(\mu_{h2},\mu)
=&~ \gamma_{hl}(\alpha_s)\,U_2(\mu_{h2},\mu)\,,
\end{align}
with initial conditions $U_1(E_\gamma,\mu_{h1},\mu_{h1})=1$ and $U_2(\mu_{h2},\mu_{h2})=1$.
The anomalous dimensions $\Gamma_{\rm cusp}$, $\gamma_H$ and $\gamma_{hl}$ are collected in Appendix~\ref{App:AD}.

For later convenience, we first investigate the RG equation for a general function $U$, which depends on anomalous dimensions $\Gamma_{\rm cusp}$ and $\gamma_{\rm gen}$,
\begin{align}
\frac{d}{d\ln\mu}\,U(E_h,\mu_{i},\mu)= &~
\Big[\Gamma_{\rm cusp}(\alpha_s)\,\ln\frac{E_h}{\mu}
+ \gamma_{\rm gen}(\alpha_s)\Big]\,
U(E_h,\mu_{i},\mu)\,,
\label{eq:gen.RG}
\end{align}
with initial condition $U(E_h,\mu_i,\mu_{i})=1$.
The solution to the general RG equation is collected in Appendix~\ref{App:AD}.
The scale $\mu$ is the factorization scale and scale $\mu_i$ is an initial scale which could be a hard scale of $\mathcal{O}(m_b)$ or a soft scale of $\mathcal{O}(\Lambda_{\rm QCD})$ or another scale.
Taking certain $E_h$ and anomalous dimensions, this equation can reproduce the RG equation satisfied by the evolution factor $U_1$ or $U_2$.
Then the two evolution factors can be obtained specifically
\begin{subequations}
\begin{align}
U_1(E_\gamma,\mu_{h1},\mu)=&~ U(2E_\gamma,\mu_{h1},\mu)
\Big|_{\gamma_{\rm gen}\to \gamma_H}\,,
\label{eq:RG.u1}
\\
U_2(\mu_{h2},\mu)=&~ U(\mu_{h2},\mu_{h2},\mu)
\Big|_{\Gamma_{\rm cusp}\to 0,\,
\gamma_{\rm gen}\to \gamma_{hl}}\,.
\label{eq:RG.u2}
\end{align}
\end{subequations}
Note that $U_2$ is independent of the cusp anomalous dimension.
Then $U_2$ can be derived from $U$ by setting $\Gamma_{\rm cusp}$ to 0 and choosing any nonzero value for $E_h$, which is chosen to be $\mu_{h2}$ here.

\subsubsection{Partial NNLL resummation for $\phi_+$}

In the Laplace space, it is relatively straightforward to solve the RG equation of the LCDA, which is~\cite{Galda:2020epp}
\begin{align}
&~ \left(\frac{d}{d\ln\mu}+\Gamma_{\rm cusp}(\alpha_s)\frac{\partial}{\partial\eta} \right)
\tilde{\phi}_+(\eta,\widehat\omega,\mu) \nonumber \\
=&~ \left\{\Gamma_{\rm cusp}(\alpha_s)\left[
\ln\frac{\widehat{\omega}}{\mu}
+\tilde{h}(-\eta)
-\psi(1+\eta) -\psi(1-\eta)
-2\gamma_E \right]-
\gamma(\alpha_s)\right\}\tilde{\phi}_+(\eta,\widehat\omega,\mu)\,,
\label{eq:RG.DA}
\end{align}
where $\psi$ is the digamma function.
The two-loop anomalous dimension $\gamma$ and the two-loop level $\tilde h$  were calculated in Ref.~\cite{Braun:2019wyx}, see the Appendix~\ref{App:AD} for details.
The explicit solution to the RG equation \eqref{eq:RG.DA} at $\alpha_s$ accuracy can be found in Ref.~\cite{Galda:2020epp}.
We would like to derive the formal $\alpha^2_s$-level solution, which could be employed to fulfill the NNLL resummation to the LCDA once the three-loop anomalous dimensions of $\gamma$ and $\tilde h$ are known.
Since any combination of $\eta+a_\Gamma(\mu_s)$ is a solution to \eqref{eq:RG.DA} with the right-hand-side terms setting to zero,
we separate the RG-evolution factor into $\eta$-independent $U_{\phi1}$ and $\eta$-dependent $U_{\phi2}$ parts, where
\begin{align}
a_{\Gamma}(\mu_s)=
-\int_{\alpha_s(\mu_s)}^{\alpha_s(\mu)}d\alpha\,
\frac{\Gamma_{\rm cusp}(\alpha)}{\beta(\alpha)}
\end{align} with $\beta$ the QCD beta function.
And the combination
$U_{\phi1}(\widehat\omega,\mu_s,\mu)\,
U_{\phi2}(\eta,\mu_s,\mu)\,
\tilde{\phi}_+(\eta+a_\Gamma(\mu_s),\widehat\omega,\mu_s)$
will be the solution to the RG equation satisfied by $\tilde{\phi}_+$.

The factor $U_{\phi1}$ satisfies the RG equation
\begin{align}
\frac{d}{d\ln\mu}\,
U_{\phi1}(\widehat\omega,\mu_s,\mu)
=&~ \left\{
\Gamma_{\rm cusp}(\alpha_s) \, \ln\frac{\widehat{\omega}}{\mu}
- \Big[ 2\gamma_E \,\Gamma_{\rm cusp}(\alpha_s) +
\gamma(\alpha_s) \Big]
\right\}
U_{\phi1}(\widehat\omega,\mu_s,\mu) \,,
\end{align}
with initial condition $U_{\phi1}(\widehat\omega,\mu_s,\mu_s)=1$.
The equation is closely related to the general RG equation \eqref{eq:gen.RG}, and we have
\begin{align}
U_{\phi1}(\widehat\omega,\mu_s,\mu)=
U(\widehat\omega,\mu_{s},\mu)
\Big|_{\gamma_{\rm gen}\to
- (\gamma+2 \gamma_E \Gamma_{\rm cusp})} \,.
\end{align}
The factor $U_{\phi2}$ is expressed as
\begin{align}
U_{\phi2}(\eta,\mu_s,\mu)
=&~
\exp\Big[
\int\limits_{\alpha_s(\mu_s)}^{\alpha_s(\mu)}
\frac{d\alpha}{\beta(\alpha)}\,
\Gamma_{\rm cusp}(\alpha) \,
\tilde{h}(\alpha,-\eta-a_\Gamma(\mu_\alpha))
\Big]\,\frac{\Gamma\big(1+\eta+a_\Gamma(\mu_s)\big)\,
\Gamma(1-\eta)}{\Gamma\big(1-\eta-a_\Gamma(\mu_s)\big)\,\Gamma(1+\eta)}
\,,
\end{align}
with $\mu_{\alpha}$ satisfing $\alpha_{s}(\mu_{\alpha})\equiv \alpha$.

Having the evolution factors $U_{\phi1}$ and $U_{\phi2}$ at hand, the large logarithms appearing in the LCDA $\tilde{\phi}_+$ can be resummed directly.
The evolution function $\tilde{\phi}_{+,U}$ is then
\begin{align}
\tilde{\phi}_{+,U}(\eta,\mu_s,\mu)=&~
\bigg( \frac{2E_\gamma\,\widehat\omega}{\mu^2} \bigg)^\eta \,
U_{\phi1}(\widehat\omega,\mu_s,\mu)\,
U_{\phi2}(-\eta,\mu_s,\mu)\,
\tilde{\phi}_+(-\eta+a_\Gamma(\mu_s),\widehat\omega,\mu_s)\,.
\label{eq:RG.phi}
\end{align}
We add the factor $(2E_\gamma\,\widehat\omega/\mu^2)^{\eta}$ into the definition of $\tilde{\phi}_{+,U}$ to ensure that it is independent of the reference parameter $\widehat\omega$.
To expand the evolution function $\tilde{\phi}_{+,U}$ in the strong coupling constant, the functions depending on $a_\Gamma(\mu_s)$ should be expanded in $\alpha_s(\mu_s)/(4\pi)$.
These expansion coefficients are collected in Appendix~\ref{App:AD}.
The NNLL resummation of the LCDA could be performed if the three-loop anomalous dimensions of $\gamma$ and $\tilde h$ are available.
While in this paper, we will set the three-loop anomalous dimensions of $\gamma$ and $\tilde h$ to zero.

Plugging the evolution factors $U_{1}$ (\eqref{eq:RG.u1}), $U_{2}$ (\eqref{eq:RG.u2}) and $\tilde{\phi}_{+,U}$ (\eqref{eq:RG.phi}) into \eqref{eq:RG.r}, we obtain the partial NNLL-resummation improved LP function $R(E_{\gamma})$.
The complete NNLL-resummation factors $U_1(E_\gamma,\mu_{h1},\mu)$ and $U_2(\mu_{h2},\mu)$ correspond to the evolution of the hard function from a hard scale $\mu_{h1}$ to the factorization scale $\mu$ and the evolution of $\tilde{f}_B$ from a hard scale $\mu_{h2}$ to $\mu$, respectively.
The factor $\tilde{\phi}_{+,U}(\eta,\mu_s,\mu)$ represents the partial NNLL evolution of the $B$-meson LCDA from a soft scale $\mu_s$ to $\mu$.

\section{Subleading-power contributions}\label{sec:NLP}

In this section, we will proceed to show the NLP corrections originated from various sources
\begin{eqnarray}\label{FVANLP:total}
\xi(E_{\gamma})
&=& \xi^{\rm fac}(E_{\gamma})
+ \xi^{\rm hp}(E_{\gamma}) \,,
\nonumber \\
\Delta \xi(E_{\gamma})
&=&  \Delta \xi^{\rm fac}(E_{\gamma})
+ \Delta\xi^{\rm hp}(E_{\gamma})\,.
\end{eqnarray}
The superscript ``hp'' stands for non-factorizable contributions from the hadronic structure of the photon, which suffer from end-point singularities.
The LO factorizable contributions calculated in this work, represented by the superscript ``fac'', can be expressed as
\begin{eqnarray}
\xi^{\rm fac}(E_{\gamma})
&=& \xi^{\rm fac}_{\rm ht}(E_{\gamma})
+ \xi^{\rm fac}_{\rm hqe}(E_{\gamma})
+ \xi^{\rm fac}_{\rm hc}(E_{\gamma})  \,,
\nonumber \\
\Delta \xi^{\rm fac}(E_{\gamma})
&=&  \Delta \xi^{\rm fac}_{\rm b}(E_{\gamma})
+ \Delta\xi^{\rm fac}_{\rm hc}(E_{\gamma}) \,.
\end{eqnarray}
The subscripts ``ht'', ``hqe'', ``b'', and ``hc'' correspond to  contributions from higher-twist $B$-meson LCDAs up to the twist six, subleading terms in the heavy-quark expansion of the $b$-quark field, the photon emitted off the $b$ quark, and subleading terms in the hard-collinear propagator, respectively.
The contribution arising from the photon emitted off the $b$ quark breaks the  large-recoil symmetry relation between the form factors.
The heavy-quark-expansion and the higher-twist-LCDA mechanisms only generate symmetry-preserving contributions at LO.
The factorizable symmetry-breaking terms are well-known~\cite{Beneke:2011nf}
\begin{align}
& \Delta \xi^{\rm fac}_{\rm b}(E_{\gamma}) =
\frac{Q_b\,\tilde{f}_B\,m_B}{2\,m_b\,E_\gamma}\,,
&&
\Delta \xi^{\rm fac}_{\rm hc}(E_{\gamma}) =
\frac{Q_u\,\tilde{f}_B\,m_B}{(2\,E_\gamma)^2}\,.
\end{align}
Below we calculate the other contributions $\xi^{\rm fac}_{\rm ht}$ \eqref{FVANLP:ht}, $\xi^{\rm fac}_{\rm hqe}$ \eqref{FVANLP:hqe}, $\xi^{\rm fac}_{\rm hc}$ \eqref{FVANLP:hc}, $\xi^{\rm hp}$ \eqref{FVANLP:hp}, and $\Delta\xi^{\rm hp}$ \eqref{FVANLP:hp}.

\subsection{The $\xi^{\rm fac}_{\rm ht}$ and $\xi^{\rm fac}_{\rm hqe}$ terms}

The higher-twist LCDA contribution can be calculated with the help of the light-cone expansion of the quark propagator
in the background gluon field~\cite{Balitsky:1987bk}
\begin{eqnarray}
\langle 0 | {\rm T} \, \{\bar q (x), q(0) \} | 0\rangle
\supset   i  g_s
\int  {d^4 \ell \over (2 \pi)^4}
{e^{- i \, \ell \cdot x} \over \ell^2 + i 0} \,
\int_0^1  d u
\left  [ u \, x_{\mu}  \gamma_{\nu}
- \frac{\slashed \ell \,\, \sigma_{\mu \nu}}
{2  \left (\ell^2 + i 0 \right )}  \right ]
 G^{\mu \nu}(u \, x) \,,
\end{eqnarray}
where $G_{\mu\nu}=i/g_s[D_\mu,D_\nu]$ with $D_{\mu} = \partial_\mu - i g_s A_\mu$.
The corresponding contribution is~\cite{Beneke:2018wjp}
\begin{eqnarray}\label{FVANLP:ht}
\xi^{\rm fac}_{\rm ht}(E_\gamma)&=&
\frac{Q_u\,\tilde{f}_B\,m_B}{(2E_\gamma)^2}
\Big[ \int_0^\infty d\omega\int_0^\infty d\xi\int_0^1du\,
\frac{\psi_4(\omega,\xi) -\tilde{\psi}_4(\omega,\xi)}
{(\omega+u\,\xi)^2}
\nonumber\\
&&-4\int_0^\infty d\omega \, \frac{\hat{g}_+(\omega)}{\omega^2} \Big]\,.
\end{eqnarray}
The $B$-meson LCDAs for $\psi_4(\omega,\xi) ,\tilde{\psi}_4(\omega,\xi)$ and $\hat{g}_+(\omega)$ are defined in Appendix~\ref{DAofB}.

Taking into account subleading terms in the matching of the $b$ quark from QCD onto HQET, the heavy-to-light current at the tree level is
\begin{eqnarray}\label{HQE}
\bar{q}\,\gamma_\mu(1-\gamma_5)\,b=
\bar{q}\,\gamma_\mu\,(1-\gamma_5)\,h_v
+ \frac{1}{2m_b}\,\bar{q}\,\gamma_\mu\,(1-\gamma_5)\,
i\not\!\! D_{\rm T}\,h_v
+\mathcal{O}(\frac{1}{m_b^2})\,,
\end{eqnarray}
where $D_{\rm T,\mu}=D_\mu-v\cdot D\, v_\mu$.
Calculating the NLP contribution from the second term on the right-hand side of the equation, we meet with the matrix element of the operator $\bar{q}(x) \,\Gamma\,D_{\rho}\,h_{v}(0)$, which can be handled by using the operator relation
\begin{align}
\bar{q}(x) \,[x,0]\,\Gamma\,D_{\rho}\,h_{v}(0)=&~
\partial_{\rho}\,\Big[\bar{q}(x)\,[x,0]\,\Gamma \,
h_{v}(0)\Big]
-\Big[\frac{\partial}{\partial x^{\rho}}\bar{q}(x)\Big] \,
[x,0]\,\Gamma \, h_{v}(0)
\nonumber \\
&+i\int^{1}_{0} du \,\bar{u}\,\bar{q}(x)\,[x,ux]\,x^{\lambda}\,
G_{\lambda\rho}(ux)\, [ux,0]\,\Gamma \,h_{v}(0)\,,
\label{eq:eom1}
\end{align}
with the Wilson line
$[z_2,z_1] = {\rm P} \, \exp\Big[ig_s \int^{z_2}_{z_1}
dz^\mu A_\mu(z)\Big]$ ensuring the gauge invariance of the operators.
The first term and the third term on the right-hand side, which correspond to two-particle and three-particle LCDA contributions, respectively, can be computed straightforwardly.
While the second term on the right-hand side of the equation provides us with troublesome operators $[\not\!\partial\bar{q}(x)]\,
\Gamma \, h_{v}(0)$ and $[n\cdot\partial\,\bar{q}(x)]\,
\Gamma \, h_{v}(0)$,
whose matrix elements are not well defined since the quark fields are not separated by a light-like distance.
These operators can be translated to desirable light-ray operators by applying operator identities
\begin{subequations}
\begin{align}
\big[\frac{\partial}{\partial x^{\rho}}\bar{q}(x)\big]\,
\gamma^\rho \, [x,0]\,
\Gamma \, h_{v}(0)=&
-i\int^{1}_{0} du \,u\,\bar{q}(x)\,[x,ux]\,x^{\lambda}\,
G_{\lambda\rho}(ux)\,\gamma^\rho \,
[ux,0]\,\Gamma \,h_{v}(0)\,,
\label{eq:eom2}  \\
\big[\frac{\partial}{\partial x^{\rho}}\bar{q}(x)\big] \,v^\rho\,
[x,0]\,\Gamma \, h_{v}(0)
=&~ v\cdot \partial \,\big[\bar{q}(x)\,[x,0]\,\Gamma \,
h_{v}(0)\big]
\nonumber\\
&+i\int^{1}_{0} du \,\bar{u}\,\bar{q}(x)\,[x,ux]\,x^{\lambda}\,
G_{\lambda\rho}(ux)\, v^\rho\, [ux,0]\,\Gamma \,h_{v}(0)\,,
\label{eq:eom3}
\end{align}
\end{subequations}
which are derived from equation of motions $\not\!\!D \,q(x)=0$ and $v\cdot D\, h_v=0$.
Then the rest of the calculations are relatively easy, and interested readers could refer to Ref.~\cite{Shen:2020hfq} for details.
Finally, one obtains for the $b$-quark expansion contribution~\cite{Beneke:2018wjp}
\begin{eqnarray}\label{FVANLP:hqe}
\xi^{\rm fac}_{\rm hqe}(E_\gamma)&=&~
\frac{Q_u\,\tilde{f}_B\,m_B}{4\,m_b\,E_\gamma}\,
\bigg\{ \frac{\bar{\Lambda}}{\lambda_B} +\int_0^\infty d\omega \,\Big[ \int_0^\omega d\eta \, \frac{\phi_+(\eta)-\phi_-(\eta)}{\omega} -\phi_+(\omega)  \Big]
\nonumber\\
&+&2\int_0^\infty d\omega \int_0^\infty \frac{d\xi}{\xi}\,\Big[ \frac{1}{\omega} +\frac{1}{\xi}\,\ln\frac{\omega}{\omega+\xi}\Big]\, \phi_3(\omega,\xi) \bigg\} \,,
\end{eqnarray}
where $\bar{\Lambda}=m_B-m_b$ is the ``effective mass" of the $B$ meson in the HQET and the inverse moment $\lambda_B$ will be introduced later in Section~\ref{sec:Numrical}.

The contributions $\xi^{\rm fac}_{\rm ht}$ and $\xi^{\rm fac}_{\rm hqe}$ can also be extracted from the $B\to\gamma\gamma$ decay~\cite{Shen:2020hfq,Qin:2022rlk}.
The $P_7$ operator in the effective Hamiltonian, which contributes to the $B\to\gamma\gamma$ decay at LP, is
\begin{align}
P_7 =
\Big(- \frac{e}{16\pi^2} \, E_\gamma \, m_b \,A^{\rm em}_\mu\Big) \,
[\bar{q} \, \gamma_\perp^\mu\, (1-\gamma_5) \not\! n \,b] \,,
\end{align}
with $A^{\rm em}$ the photon field.
If the operator $[\bar{q} \, \gamma_\perp^\mu\, (1-\gamma_5) \not\! n \,h_{v}]$ can be simplified to $2 [\bar{q} \, \gamma_\perp^\mu\, (1-\gamma_5)\,h_{v}]$ which is the heavy-to-light current in the $\decay$ decay, we can just get the corresponding contribution to the $\decay$ decay from the $B\to\gamma\gamma$ result.
This is actually the case for $\xi^{\rm fac}_{\rm ht}$ and $\xi^{\rm fac}_{\rm hqe}$ contributions at the tree level, because the hard-collinear propagator provides us with the $(\cdots)\not\!\bar{n}$ structure which turns $\not\! n \,h_{v}$ to $2\not\!v \,h_{v}=2h_{v}$.
%{\color{red} 
The contributions $\xi^{\rm fac}_{\rm ht}$ and $\xi^{\rm fac}_{\rm hqe}$ are already presented in~\cite{Beneke:2018wjp}, they can also be extracted from the $B \to\gamma\gamma$ decay.
%}

\subsection{Contribution from the hard-collinear propagator}

Subleading terms in the hard-collinear propagator will contribute to the decay form factors at NLP~\cite{Gao:2019lta,Wang:2021yrr,Gao:2021sav,Cui:2022zwm,Cui:2023jiw}.
Employing $k^2=0$ with $k$ the momentum of the light quark in the $B$ meson, the hard-collinear propagator reads
\begin{eqnarray}
\frac{i\,(\slash{p}-\slash{k})}{(p-k)^2}=
-\frac{i}{\bar{n}\cdot k}\, \frac{\not\!\bar{n}}{2}
+\Big[
\frac{i \, n\cdot k}{2\, E_\gamma\,\bar{n}\cdot k}\,\frac{\not\!\bar{n}}{2}
+\frac{i \not\!k_\perp}{2\, E_\gamma\,\bar{n}\cdot k}
+\frac{i}{2\, E_\gamma}\,\frac{\not\! n}{2}
\Big]\,.
\label{eq:hc.expan}
\end{eqnarray}
The first term on the right-hand side is the LP term.
The terms in the square brackets generate NLP contributions
with the last term contributing to $\Delta\xi^{\rm fac}_{\rm hc}$.
The first two terms in the square brackets provide us with a non-local symmetry-preserving contribution at LO
\begin{eqnarray}\label{FVANLP:hc}
\xi^{\rm fac}_{\rm hc}(E_\gamma)&=&~\frac{Q_u\,\tilde{f}_B\,m_B}{2\,E_\gamma^2}\, \Big[\int_0^\infty d\omega\int_0^\infty\frac{d\xi}{\xi}\, \Big(\frac{1}{\xi}\,\ln\frac{\omega}{\omega+\xi}+\frac{1}{\omega}\Big)\, \psi_4(\omega,\xi)-\frac{\bar{\Lambda}}{\lambda_B}+\frac{1}{2}\Big]\,.
\end{eqnarray}
As in the calculation of the contribution $\xi^{\rm fac}_{\rm hqe}$, we have used the operator identities in \eqref{eq:eom3} to compute $\xi^{\rm fac}_{\rm hc}$.

The symmetry-preserving contribution $\xi^{\rm fac}_{\rm hc}$ can also be obtained from results in the $B\to\gamma\gamma$ decay~\cite{Shen:2020hfq}.
Taking a closer look at terms in \eqref{eq:hc.expan}, we find that the $\not\!k_\perp$ term does not contribute to the $\decay$ and the $B\to\gamma\gamma$ decays at LO.
For the contribution from the $n\cdot k$ term, the operator $P_7$ can be simplified to the heavy-to-light current $[\bar{q} \, \gamma_\perp^\mu\, (1-\gamma_5)\,b]$.
Then we conclude that the non-local symmetry-preserving contributions in the two processes are the same up to a trivial pre-factor.

\subsection{Contribution from the resolved photon}
%{\color{red} 
Except for the above direct-photon contributions there is contribution from the resolved photon.
After integrating out the hard-collinear scale effects, the SCET$_{\rm II}$ operators for the first kind of contributions will include a collinear photon field $A^{\rm em}_{c}$.
While for the resolved-photon contribution, the photon is represented by higher collinear Fock states, for example, two collinear quark fields $[\bar \xi \, \xi]_c$.
Contracting the photon field with the photon state, the direct-photon contributions are obtained by introducing the electromagnetic current $j^\nu_{\rm em}$ and calculating the corresponding vacuum-to-$B$-meson correlation function in \eqref{eq:correlation.function}.
The non-perturbative inputs corresponding for the matrix elements of soft operators are the $B$-meson LCDAs.
For the resolved-photon contribution, we can not introduce the electromagnetic current and will have to keep the photon state. Thus apart from the $B$-meson LCDA, the LCDAs of the photon are also necessary.

As suggested in~[13] we will calculate this resolved photon correction directly with QCD method.
%}
The subleading SCET$_{\rm II}$ operator
${\cal O} \supset [\bar q_s \, h_v]_s \,\, [\bar \xi \, \xi]_c$,
which contains collinear fields, provides power-suppressed contribution to the $\decay$ decay~\cite{Beneke:2003pa}.
The collinear part of its matrix element
$\langle \gamma(p) | [\bar \xi \, \xi]_c | 0 \rangle $, which defines the photon LCDA, represents the contribution from the hadronic structure of the photon.
While end-point divergences will appear when trying to derive the factorization formula for this contribution.
In this paper we employ the LCSR with photon LCDA to calculate the resolved-photon contribution.

The photon-to-vacuum correlation function with an interpolating
current (with momentum $p+q$) for the $B$ meson is constructed as
\begin{eqnarray}\label{CorrelationFun:HadronicPhoton}
\Pi^{\rm hp}_\mu(p,q)=\int d^4x\,e^{iq\cdot x}\,\langle\, \gamma(p)\,|\,T\{ \bar{u}(x)\,\gamma_{\mu\perp}\,(1-\gamma_5)\,b(x), \; \bar{b}(0)\, \gamma_5\,u(0) \}\,|0\,\rangle\,,
\end{eqnarray}
with $q^2<m^2_b$.
We first calculate the correlation function in the region $(p+q)^2<m^2_b$ and then apply the dispersion relation to analytically continue the result to the physical region.
We also employ the power-counting schemes $(p+q)^2-m^2_b \sim \mathcal{O}(m^2_b)$ to avoid the end-point divergence and $q^2-m^2_b \sim \mathcal{O}(m^2_b)$ inherited from $E_\gamma\sim\mathcal{O}(m_b)$.

First, we consider the leading twist-2 photon LCDA contribution to the form factors.
The hard-collinear factorization formula for the correlation function is
\begin{align}
\Pi^{\rm hp}_{\mu,\rm LT}(p,q)=
-Q_u\, g_{\rm em}\, n\cdot p\, \chi(\nu)\,
\langle \bar{q}q\rangle(\nu)\,\varepsilon^{*\alpha}\,
(g_{\mu\alpha}^\perp-i\epsilon_{\mu\alpha nv})
\int_0^1 du\,\phi_\gamma(u,\nu)\,
H_\gamma(p,q,u,\nu)\,,
\end{align}
with $\chi$ the magnetic susceptibility of the quark condensate, $\langle \bar{q}q\rangle$ the quark condensate, $\varepsilon^\alpha$ the polarization vector of the photon, and $u$ the momentum fraction carried by the quark in the photon.
Here the scale $\nu$ stands for the factorization scale for the photon-to-vacuum correlation function.
The long-distance photon LCDA $\phi_\gamma$ is defined in Appendix~\ref{DAofPhoton}.
The hard matching coefficient $H_\gamma$ is recalculated up to NLO with NDR scheme of $\gamma_5$~\cite{Li:2020rcg}
\begin{eqnarray}
H_\gamma^{(0)}(p,q,u,\nu)
&=&\frac{i}{2}\,\frac{\bar{n} \cdot q}{u(p+q)^2+\bar{u}\,q^2-m_b^2+i0}
\\
H_\gamma^{(1)} (p,q,u,\nu)
& =& H_\gamma^{(0)}(p,q,u,\nu)\, \bigg \{ -2 \,
\bigg[ {\overline{r}_{2} \over r_1-r_2} \,
\ln {\overline{r}_{1} \over \overline{r}_{2}}
+ {\overline{r}_{3} \over r_1-r_3}  \,
\ln {\overline{r}_{1} \over \overline{r}_{3}}
+ {3 \over\overline{r}_{1}}\bigg]\,
\ln {\nu^2 \over m_b^2}
\nonumber \\
&&+ 2\,\bigg[\Big(\frac{\overline{r}_{2}}{r_1-r_2}
+\frac{\overline{r}_{3}}{r_1-r_3}\Big)\,
\Big({\rm Li}_2(r_1)+\ln^2\overline{r}_{1} \Big)
-\frac{\overline{r}_{2}}{r_1-r_2}\,
\Big({\rm Li}_2(r_2)+\ln^2\overline{r}_{2} \Big)
\nonumber\\
&&-\frac{\overline{r}_{3}}{r_1-r_3}\,
\Big({\rm Li}_2(r_3)+\ln^2\overline{r}_{3} \Big) \bigg]
+\bigg[\frac{\overline{r}_{1}}{r_1}\Big(\frac{1-3r_1}{r_1-r_2}-\frac{2}{r_1-r_3}\Big)\,
+\frac{\overline{r}^{2}_{1}}{r_1^2}-4\bigg]\,\ln\overline{r}_{1}
\nonumber\\
&&
-\frac{\overline{r}_{2}\,(1-3\,r_2)}{r_2\,(r_1-r_2)}\,
\ln\overline{r}_{2}
+\frac{2\,\overline{r}_{3}}{r_3\,(r_1-r_3)}\,\ln\overline{r}_{3}
+\frac{1}{r_1} -\frac{8}{\overline{r}_{1}} -3 \bigg \}\,,
\label{one-loop QCD amplitude at twist-2}
\end{eqnarray}
where $r_1=(u\,p+q)^2/m_b^2$, $r_2=q^2/m_b^2$, $r_3=(p+q)^2/m_b^2$, and $\overline{r}_{i}\equiv1-r_{i}$.
Choosing the factorization scale $\nu$ of $\mathcal{O}(m_b)$, the combined factor $[\chi(\nu) \langle \bar{q}q\rangle(\nu) \phi_\gamma(u,\nu)]$ need to be evolved from a non-perturbative scale $\nu_{0}$ to $\nu$ by applying the RG-equation approach, see Ref.~\cite{Wang:2018wfj} for details.

Employing the standard LCSR approach, which includes applications of the dispersion relation, the quark-hadron duality, and the Borel transformation, the NLL-resummation improved resolved-photon contribution with twist-2 photon LCDA is
\begin{eqnarray}
\xi^{\rm hp}_{\rm LT}(E_\gamma) &=&
Q_u \, \frac{m_b+m_u}{2\, f_B} \,
 \chi(\nu) \, \langle \bar q q \rangle(\nu)
\int_{0}^{s_0} \, ds \, {\rm exp}
\left [ - {s -m_B^2 \over M^2} \right ] \,\rho(s, q^2,\nu) \,.
\end{eqnarray}
Here $M^2$ is the Borel mass and $s_0$ represents the integral threshold.
In the sum rules, $\xi^{\rm hp}_{\rm LT}$ is independent of $(p+q)^2$ but still depends on $q^2$, thus the sum rules is valid only in the region $q^2<m^2_b$ (or $E_\gamma>(m^2_B-m^2_b)/(2m_B)$).
The newly derived spectral density is written as
\begin{eqnarray}
\rho(s, q^2,\nu)= \rho^{(0)}(s, q^2,\nu)
+ \frac{\alpha_s(\nu) \, C_F}{4\pi} \,\rho^{(1)}(s, q^2,\nu)\,,
\end{eqnarray}
where
\begin{eqnarray}
m_b^2\,\rho^{(0)}(s, q^2,\nu)&=&
-\frac{1}{w_{sr}}\,\phi_\gamma
\Big(\frac{\overline{r}_{2}}{w_{sr}},\nu \Big)\,\theta(\hat{s})\,,
\\
m_b^2\,\rho^{(1)}(s, q^2,\nu)
&=&- {2 \over w_{sr}} \,\ln{\nu^2 \over m_b^2}\,
\bigg\{
\, \int_0^1 \, d u \, \Big[ {1-2u \over u\bar{u}}\,\theta(u_{sr})
+{1\over \bar u}\,\theta(\hat{s})\,\Big] \,
\phi_{\gamma}(u, \nu)
\nonumber \\
&& \hspace{-1.5cm} +
\int_0^1 \, d u \,
\bigg[ \ln{u^2_{sr} \over \hat{s} \, \overline{r}_{2}}\,
\theta(u_{sr})
-\, \ln \bigg |{ u_{sr} \over \hat{s}} \bigg | \,
\theta(\hat{s}) \bigg] \,
\phi_{\gamma}^{\prime}(u, \nu)
 + \, {3 \over w_{sr}} \, \theta(\hat{s})\,
 \phi_{\gamma}^{\prime}
\Big({\overline{r}_{2} \over w_{sr}}, \nu \Big)\, \bigg\}
\nonumber\\
&& \hspace{-1.5cm} +
\int_0^1du\,\phi_\gamma(u,\nu) \,\bigg\{
\theta(u_{sr}) \bigg[
\frac{2\,(1-2u)}{u\,\bar{u}\,w_{sr}}\, \ln\frac{u^{2}_{sr}}{u_{sr}+1}
+\frac{6\,\ln u_{sr}}{(u_{sr}+1)^2}
+\frac{4}{u_{sr}}\,\ln(u_{sr}+1)
\nonumber \\
&& \hspace{-1.5cm}
+\Big(\frac{2\,(2u-1)}{u\,\bar{u}\,w_{sr}}+
\frac{3\,\overline{r}_{2}}{u\,w_{sr}}+1\Big)\, \frac{1}{u_{sr}+1}
\bigg]
-2\,\theta(\hat{s})
\Big(\frac{1}{\hat{s}+1}+\ln \frac{\hat{s}+1}{\hat{s}^{2}}\Big)
\Big({\cal P} \frac{1}{u_{sr}}+ \frac{1}{\bar{u}\,w_{sr}}\Big)
\bigg\}
\nonumber \\
&& \hspace{-1.5cm}
+ {1 \over w_{sr}}\int_0^1 \, d u \,
\phi_{\gamma}^{\prime}(u, \nu) \, \theta(u_{sr})  \,
\bigg[4 \,
\ln^2u_{sr}
+ \ln\frac{u^2_{sr}}{u_{sr}+1}
+\frac{1-6\ln u_{sr}}{u_{sr}+1}
- {4\over 3}\,\pi^2\bigg]
\nonumber \\
&& \hspace{-1.5cm}
+\frac{2}{w_{sr}}\,
\bigg[{\rm Li}_2(1-\overline{r}_{2}) -{\rm Li}_2(-\hat{s})-\ln (\hat{s}+1)\,\ln\hat{s}+ \ln^2 \overline{r}_{2} + \ln^2 \hat{s}
+\frac{3\,\overline{r}_{2}-2}{2(1-\overline{r}_{2})}\,\ln\overline{r}_{2}
\nonumber \\
&& \hspace{-1.5cm}
-\frac{\ln\hat{s}}{\hat{s}+1}
+\frac{3}{2}-\frac{7}{6}\pi^2 \bigg]\,
\theta(\hat{s})\,
\phi_\gamma\Big(\frac{\overline{r}_{2}}{w_{sr}},\nu\Big)
-8\,
\frac{1}{w_{sr}^2}\,
\theta(\hat{s})\,
\phi'_\gamma\Big(\frac{\overline{r}_{2}}{w_{sr}},\nu\Big)\,,
\end{eqnarray}
with $\hat{s}=s/m^2_b-1$, $w_{sr}=\hat{s}+\overline{r}_{2}$, and $u_{sr}=u\,\hat{s}-\bar{u}\,\overline{r}_{2}$.

The two-particle and three-particle photon LCDAs up to twist four also generate NLP contributions at LO.
Note that there are a few misprints in definitions of higher-twist photon LCDAs in Ref.~\cite{Wang:2018wfj}, we therefore recalculate corrections from the higher-twist photon LCDAs which are updated in Appendix~\ref{DAofPhoton}.
The corresponding NLP contributions at the tree level are
\begin{align}
\xi^{\rm hp}_{\rm 2PHT}
=&-Q_u\,\frac{m_b+m_u}{4\,f_B\,m_B}\,\frac{1}{m_b^2-q^2}\,
\exp\Big(\frac{m_B^2}{M^2}\Big)\,
\nonumber\\
&\times \bigg\{\bigg[
\exp\Big(-\frac{s_0}{M^2}\Big)\,
\rho^{\rm a}_{\rm 2PHT}(\kappa(s_0),\nu)
+\frac{1}{M^2} \int_{m_b^2}^{s_0} ds\,
\exp\Big(-\frac{s}{M^2}\Big)\,
\rho^{\rm a}_{\rm 2PHT}(\kappa(s),\nu) \bigg]
\nonumber\\
&- \frac{1}{2}\bigg[
\exp\Big(-\frac{s_0}{M^2}\Big)\,
\Big( \frac{1}{m_b^2-q^2}
+\frac{1}{M^2\,\kappa_0}
+\frac{1}{\kappa_0}\frac{d}{d\kappa_0}\,
\Big)\rho^{\rm b}_{\rm 2PHT}(\kappa_0,\nu)
\Big|_{\kappa_0\to\kappa(s_0)}
\nonumber\\
&+ \frac{1}{M^4} \int_{m_b^2}^{s_0}ds\,
\exp\Big(-\frac{s}{M^2}\Big)\,\frac{1}{\kappa(s)}\,
\rho^{\rm b}_{\rm 2PHT}(\kappa(s),\nu) \bigg] \bigg\}\,,
\\
\xi^{\rm hp}_{\rm 3PHT}=&
-\frac{m_b+m_u}{f_B\,m_B}\,
\frac{1}{m_b^2-q^2}\,
\exp\Big(\frac{m_B^2}{M^2}\Big)\,
\langle\bar{q}q\rangle(\nu)\,
\nonumber\\
& \times \bigg\{
\exp\Big(-\frac{s_0}{M^2}\Big)
\int_0^{\kappa(s_0)}d\alpha_q
\int_{\kappa_{0q}}^{1-\alpha_q} \frac{d\alpha_g}{\alpha_g}
\Big[Q_u\,\rho^{G}_{\rm 3PHT}
+Q_b\, \rho^{\gamma}_{\rm 3PHT}\Big]\Big(\alpha_q,\alpha_g,\frac{\kappa_{0q}}{\alpha_g},\nu\Big)
\nonumber\\
&+ \frac{1}{M^2}
\int_{m_b^2}^{s_0}ds \int_0^{\kappa(s)}d\alpha_q
\int_{\kappa_{q}}^{1-\alpha_q}
\frac{d\alpha_g}{\alpha_g} \,
\exp\Big(-\frac{s}{M^2}\Big)\,
\nonumber\\
& \times \Big[Q_u\,\rho^{G}_{\rm 3PHT}
+Q_b\,\rho^{\gamma}_{\rm 3PHT}\Big]\Big(\alpha_q,\alpha_g,\frac{\kappa_{0q}}{\alpha_g},\nu\Big)
\Big|_{\kappa_{0q}\to\kappa(s_0)-\alpha_q,\,
\kappa_{q}\to\kappa(s)-\alpha_q} \bigg\}\,,
\\
\Delta\xi^{\rm hp}_{\rm 2PHT}
=&~ \xi^{\rm hp}_{\rm 2PHT}
\Big|_{\scriptsize
\begin{array}{l}
\rho^{\rm a}_{\rm 2PHT}\to\Delta\rho^{\rm a}_{\rm 2PHT} \\
\rho^{\rm b}_{\rm 2PHT}\to\Delta\rho^{\rm b}_{\rm 2PHT}
\end{array}
} \,,
\qquad
\Delta\xi^{\rm hp}_{\rm 3PHT}=
\xi^{\rm hp}_{\rm 3PHT}
\Big|_{\scriptsize
\begin{array}{l}
\rho^{G}_{\rm 3PHT}\to\Delta\rho^{G}_{\rm 3PHT} \\
\rho^{\gamma}_{\rm 3PHT}\to\Delta\rho^{\gamma}_{\rm 3PHT}
\end{array}
} \,,
\end{align}
where $\kappa(s)=(m_b^2-q^2)/(s-q^2)$, and the spectral densities are
\begin{eqnarray}
\rho^{\rm a}_{\rm 2PHT}(u,\nu)&=&
m_b\,f_{3\gamma}(\nu)\,\Big[\psi^{(a)}
-2\,\bar{\psi}^{(v)}\Big](u,\nu)
- \langle\bar{q}q\rangle(\nu)\,
\Big[\mathbb{A} + \bar{h}_\gamma\Big](u,\nu)\,,
\nonumber\\
\Delta\rho^{\rm a}_{\rm 2PHT}(u,\nu)&=&
m_b\,f_{3\gamma}(\nu)\,\Big[\psi^{(a)}
+2\,\bar{\psi}^{(v)}\Big](u,\nu)
+ \langle\bar{q}q\rangle(\nu)\,\bar{h}_\gamma(u,\nu)\,,
\nonumber\\
\rho^{\rm b}_{\rm 2PHT}(u,\nu)&=&
2\,m_b^2\,\langle\bar{q}q\rangle(\nu)\,
\Big[\mathbb{A}+\bar{h}_\gamma\Big](u,\nu)\,,
\nonumber\\
\Delta\rho^{\rm b}_{\rm 2PHT}(u,\nu)&=&
-2\,m_b^2\,\langle\bar{q}q\rangle(\nu)\,\bar{h}_\gamma(u,\nu)\,,
\\
\rho^{G}_{\rm 3PHT}(\alpha_q,\alpha_g,z,\nu)&=&
z\,\Big(T_1 - 2\,T_2 - T_3 + T_4 \Big)(\underline{\alpha};\nu)
+ \bar{z}\,\Big(S - \tilde{S} + T_1 + T_2 + T_3 \Big)(\underline{\alpha};\nu)
\,,
\nonumber\\
\Delta\rho^{G}_{\rm 3PHT}(\alpha_q,\alpha_g,z,\nu)&=&
z\, \Big(S + \tilde{S} + T_2\Big)(\underline{\alpha};\nu)
-\bar{z}\, \Big(2\,T_2 + T_4\Big)(\underline{\alpha};\nu)
\nonumber\\
\rho^{\gamma}_{\rm 3PHT}(\alpha_q,\alpha_g,z,\nu)&=&
\bar{z}\,S_\gamma(\underline{\alpha};\nu)
+z\, T_4^{\gamma}(\underline{\alpha};\nu) \,,
\nonumber\\
\Delta\rho^{\gamma}_{\rm 3PHT}(\alpha_q,\alpha_g,z,\nu)&=&
z \, S_\gamma(\underline{\alpha};\nu)
-\bar{z}\, T_4^{\gamma}(\underline{\alpha};\nu) \,,
\end{eqnarray}
with $\bar{z}\equiv 1-z$, $\underline{\alpha}\equiv \{\alpha_q,1-\alpha_q-\alpha_g,\alpha_g\}$ and $f_{3\gamma}$ a non-perturbative parameter.
The two-particle LCDAs $\psi^{(a)}$ and $\bar{\psi}^{(v)}$ are twist-3 LCDAs, and all the other LCDAs are of twist four.
One can then conclude that, at the tree level, the three-particle twist-3 photon LCDAs do not contribute to the $\decay$ decay at NLP.

In summary, the resolved-photon contribution is
\begin{eqnarray}\label{FVANLP:hp}
\xi^{\rm hp}(E_\gamma)&=&
\xi^{\rm hp}_{\rm LT}(E_\gamma)
+ \xi^{\rm hp}_{\rm 2PHT}(E_\gamma)
+ \xi^{\rm hp}_{\rm 3PHT}(E_\gamma) \,,
\nonumber \\
\Delta\xi^{\rm hp}(E_\gamma)&=&
\Delta\xi^{\rm hp}_{\rm 2PHT}(E_\gamma)
+ \Delta\xi^{\rm hp}_{\rm 3PHT}(E_\gamma)  \,.
\end{eqnarray}
It is clear that the leading twist-2 photon LCDA only generates symmetry-preserving contributions, while the higher-twist photon LCDAs contribute both to the symmetry-preserving and symmetry-breaking terms.
%{\color{red} 
We would like to mention that the resolved photon contribution could be extracted form the results of the charged $\bar B_u\to\gamma$ form factors in~\cite{Janowski:2021yvz}. The twist-two contributions up-to NLO are the same the twist-two ones in~\cite{Janowski:2021yvz}. 
Our twist-three and twist-four corrections at LO also reproduce the corresponding twist-three and four light-cone OPE results in~\cite{Janowski:2021yvz}.
In addition to the non-local photon LCDA contributions to the $B\to\gamma$ form factors up to the twist-4, Ref.~\cite{Janowski:2021yvz} also calculated the twist-1 and twist-4 local OPE contributions at NLO.
%}

\section{Numerical analysis}\label{sec:Numrical}

We are now ready to show the phenomenological impact of higher-order and
higher-power corrections to the $\decay$
decay. In doing that, we first specify the theoretical inputs used in this work.

\subsection{Theoretical inputs}

We employ the Mathematica package \textsf{RunDec}~\cite{Herren:2017osy} for the
five-loop running of strong coupling constant with the initial condition $\alpha_s(m_Z)=0.1181$ and $m_Z=91.1876$~GeV in the context of the number of active flavor $n_f=5$, and also adopt
the threshold values $\mu_4=5.0$ GeV and $\mu_3=1.2$ GeV for crossing
$n_f=4$ and $n_f=3$, respectively. The package \textsf{HPL}~\cite{Maitre:2005uu,Maitre:2007kp}
is used to calculate harmonic polylogarithms in the hard function.

\begin{table}[t]
\centering
\setlength\tabcolsep{4pt} \def\arraystretch{1.5}
\begin{tabular}{|llc|llc|}
\hline
  ~Parameter~ & ~Value~ & Ref. & ~Parameter~ & ~Value~ & Ref.\\
  \hline
  $f_B$ & $0.190\pm 0.0013$~GeV &\cite{Aoki:2019cca} &
  $m_b^{\rm pole}$ & $4.8\pm0.1$~GeV & \cite{Shen:2020hfq} \\
%  \hline
  $\overline{m}_b(\overline{m}_b)$ & $4.198\pm0.012$~GeV &\cite{Aoki:2019cca} &
  $\overline{m}_u(2\rm{GeV})$ & $2.32\pm0.10$~MeV &\cite{Zyla:2020zbs}\\
%  \hline
$\mu$ & $1.5\pm0.5$~GeV &
& $\mu_{h1},\mu_{h2}$ & $[\overline{m}_b/2,2\overline{m}_b]$ & \\
%  \hline
$\mu_{s}$ & 1.0~GeV && $\widehat{\omega}$ & $0.197^{+0.199}_{-0.099}$~GeV &\\
%  \hline
  $\lambda_B(\mu_s)$ & $0.350\pm0.150$~GeV &\cite{Beneke:2020fot} &
$\sigma_B^{(2)}(\mu_s)$ & $1.5\pm 1.0$ & \\
%  \hline
$\sigma_B^{(3)}(\mu_s)$ & $0.0\pm 6.0$&& $\sigma_B^{(4)}(\mu_s)$ & $0.0\pm 6.0$ & \\
%  \hline
$(\lambda_E^2/\lambda_H^2)(\mu_s)$ & $0.50\pm0.10$ &\cite{Beneke:2018wjp}   & $(2\lambda_E^2+\lambda_H^2)(\mu_s)$ &
  $0.25\pm0.15$~GeV$^{2}$ & \cite{Beneke:2018wjp} \\
%\hline
\hline
\multicolumn{6}{|c|}{Resolved-photon parameters}\\
\hline
$\nu$ & $[\overline{m}_b/2,2\overline{m}_b]$ & &
$\nu_0$ & 1.0~GeV & \\
  $\langle\bar{q}q\rangle(\nu_0)$ &
  $(-266\pm21~\rm{MeV})^3$ &\cite{Zyla:2020zbs} &
$\chi(\nu_0)$ & $3.15\pm0.3$~GeV$^{-2}$ &\cite{Ball:2002ps} \\
%  \hline
$f_{3\gamma}(\nu_0)$ & $-(4\pm2)\times 10^{-3}$~GeV$^{-2}$ &\cite{Ball:2002ps}
& $a_2(\nu_0)$ & $0.07\pm0.07$ &\cite{Ball:2002ps} \\
%  \hline
$\omega^A_{\gamma}(\nu_0)$ & $-2.1\pm1.0$ &\cite{Ball:2002ps}
& $\omega^V_{\gamma}(\nu_0)$ &
  $3.8\pm1.8$ & \cite{Ball:2002ps} \\
%  \hline
  $\kappa(\nu_0)$ & $0.2\pm0.2$&\cite{Ball:2002ps}
  & $\kappa^+(\nu_0)$ & 0 &\cite{Ball:2002ps}\\
%  \hline
$\zeta_{1}(\nu_0)$ & $0.4\pm0.4$ &\cite{Ball:2002ps}
& $\zeta_1^+(\nu_0),\,\zeta_2^+(\nu_0)$
& 0 &\cite{Ball:2002ps}\\
% \hline
$s_0$ & $37.5\pm2.5$~GeV$^2$ & \cite{Wang:2018wfj} & $M^2$ & $18.0\pm3.0$~GeV$^2$ & \cite{Wang:2018wfj} \\
  \hline
\end{tabular}
\caption{Summary of the numerical values for input parameters
implemented in the phenomenological analysis of the
$\decay$ decay.}
\label{Table:InputParameters}
\end{table}

The leading-twist $B$-meson LCDA serves as the most important non-perturbative input.
Thanks to the smallness of $a_\Gamma$, we can expand the LCDA around zero~\cite{Galda:2020epp}
\begin{align}
\tilde{\phi}_+(\eta,\widehat{\omega},\mu_s)
= \frac{1}{\lambda_B(\mu_s)} \,
\Big[
1+ \sum_{n\geq 2} \frac{\eta^n}{n!} \, \sigma_B^{(n)}(\mu_s)
\Big] \,,
\label{eq:DAexp}
\end{align}
and truncate the expansion at a large $n=4$.
The inverse(-logarithmic) moments of the $B$-meson LCDA are defined as
\begin{align}
\frac{1}{\lambda_B(\mu)}
=& \int_0^{\infty}\, \frac{d\omega}{\omega} \,
\phi_+(\omega,\mu)
= \tilde{\phi}_+(0,\widehat{\omega},\mu) \,,
\nonumber \\
\sigma_B^{(n)}(\mu)
=&~ \lambda_B(\mu)
\int_0^{\infty}\, \frac{d\omega}{\omega}\,
\ln^n\frac{\widehat{\omega}}{\omega}\, \phi_+(\omega,\mu)
= \lambda_B(\mu)\, \Big(\frac{d}{d\eta}\Big)^n
\tilde{\phi}_+(\eta,\widehat{\omega},\mu) \Big|_{\eta\to 0}\,.
\end{align}
In the expansion of $\tilde{\phi}_+$, we have choose certain value for $\widehat{\omega}$ such that $\sigma_B^{(1)}(\mu_{s})\equiv0$
\footnote{By introducing
\begin{equation}
\widehat{\sigma}_B^{(n)}(\mu)
=\lambda_B(\mu)
\int_0^{\infty}\, \frac{d\omega}{\omega}\,
\ln^n \frac{\lambda_B(\mu)e^{-\gamma_E}}{\omega}\,
\phi_+(\omega,\mu)\,,
\end{equation}
the parameter $\widehat{\omega}$ is determined though
$
\widehat{\omega}=
\lambda_B(\mu_{s})\,
\exp[-\widehat{\sigma}_B^{(1)}(\mu_s)-\gamma_E]
$.
From $\lambda_{B}(\mu_{s})=0.35$~GeV and $\widehat{\sigma}_B^{(1)}(\mu_s) = 0.0\pm0.7$, we obtain $\widehat{\omega}=0.197^{+0.199}_{-0.099}$~GeV.
}.
The values for parameters of the $B$-meson LCDA are summarized in Tab.~\ref{Table:InputParameters}.

The higher-twist $B$-meson LCDAs fulfill certain asymptotic behaviours and are also constrained by the equation of motions.
Applying general forms of the $B$-meson LCDAs proposed in Ref.~\cite{Beneke:2018wjp}, which take into consideration the above constraints, the NLP contributions involving the $B$-meson LCDAs could be remarkably simplified
\begin{align}
\xi^{\rm ht}(E_\gamma)
=& -\frac{Q_u\,\tilde{f}_B\,m_B}{2\,E_\gamma^2}\,
\Big[\frac{2\,(\lambda_E^2+2\lambda_H^2)}{6\bar{\Lambda}^2
+2\lambda_E^2+\lambda_H^2} +\frac{1}{2}\Big]\,,
\nonumber \\
\xi^{\rm hqe}(E_\gamma)
=&~ \frac{Q_u\,\tilde{f}_B\,m_B}{4\,m_{b}\,E_\gamma}\,
\Big[\frac{\bar{\Lambda}}{\lambda_{B}}-2
+\frac{4\,(\lambda_E^2-\lambda_H^2)}{6\bar{\Lambda}^2
+2\lambda_E^2+\lambda_H^2}\Big]\,,
\nonumber \\
\xi^{\rm hc}(E_\gamma)
=&~ \frac{Q_u\,\tilde{f}_B\,m_B}{2\,E_\gamma^2}\,
\Big[\frac{\lambda_E^2(12-\pi^2)}{ 6\bar{\Lambda}^2+2\lambda_E^2+\lambda_H^2}
-\frac{\bar{\Lambda}}{\lambda_B} +\frac{1}{2}\Big]\,,
\end{align}
where the factorization-scale dependence of the parameters $\tilde{f}_B$, $\lambda_B$, $\bar{\Lambda}$, and $\lambda^{2}_{E,H}$ should be understood \footnote{
The HQET parameters $\lambda^{2}_E$ and $\lambda^{2}_H$ are defined by the hadronic matrix element
of the three-body effective local operator~\cite{Grozin:1996pq,Braun:2017liq}
\begin{align}
&~ \langle 0 | \,
\bar q (0) \, g_s \, G_{\mu \nu}(0) \, \Gamma \, h_v(0) \,
| \bar B(v) \rangle
\nonumber \\
=& - {\tilde{f}_{B}(\mu) \, m_{B} \over 6} \,
{\rm Tr} \left \{ \gamma_5 \, \Gamma \,
{1 + \slashed v \over 2}\,
\Big [ \lambda_H^2(\mu)  \, \left (  i \, \sigma_{\mu \nu} \right )
+ [\lambda_H^2(\mu) - \lambda_E^2(\mu)] \,
\left ( v_{\mu} \, \gamma_{\nu} - v_{\nu} \, \gamma_{\mu} \right )
\Big ]  \right \}   \,.
\end{align}
}.
We also apply the LL resummation to $\lambda^{2}_{E,H}$ employing RG equations of them~\cite{Grozin:1996hk,Nishikawa:2011qk} (see Ref.~\cite{Shen:2020hfq} for details).
The values for $\lambda^{2}_{E,H}$ at the intrinsic scale are collected in Tab.~\ref{Table:InputParameters}, and we use the relation $\bar{\Lambda}=m_B-m^{\rm pole}_b$ to determining $\bar{\Lambda}$ with $m^{\rm pole}_b$ the pole mass of the $b$ quark.

For the sake of calculating the resolved photon corrections,
phenomenological models for the photon LCDAs have to be introduced.
The series expansion of the leading-twist photon LCDA in terms of the Gegenbauer polynomials and the conformal expansions of the chiral-even twist-three and chiral-odd
twist-four LCDAs can be found in Ref.~\cite{Wang:2018wfj}. The relevant parameters are collected in Tab.~\ref{Table:InputParameters}.

In addition, the factorization scale $\mu$ is varied in the range $\mu=(1.5\pm 0.5)$~GeV.
We also vary the hard scales $\mu_{h1}$ and $\mu_{h2}$ in the interval $[\overline{m}_b/2,2\overline{m}_b]$ around the defaulted value $\overline{m}_b$.
We adopt the $B$-meson decay constant from the lattice gauge simulations collected in the PDG~\cite{Zyla:2020zbs}.
The up and bottom quark masses are chosen in the context of
$\overline{\rm MS}$ scheme from the PDG~\cite{Zyla:2020zbs}.
For the sake of convenience, we list the parameters in Tab.~\ref{Table:InputParameters}.

\subsection{Predictions for the form factors}

\begin{figure}[!h]
  \centering
\includegraphics[width=.98\textwidth]{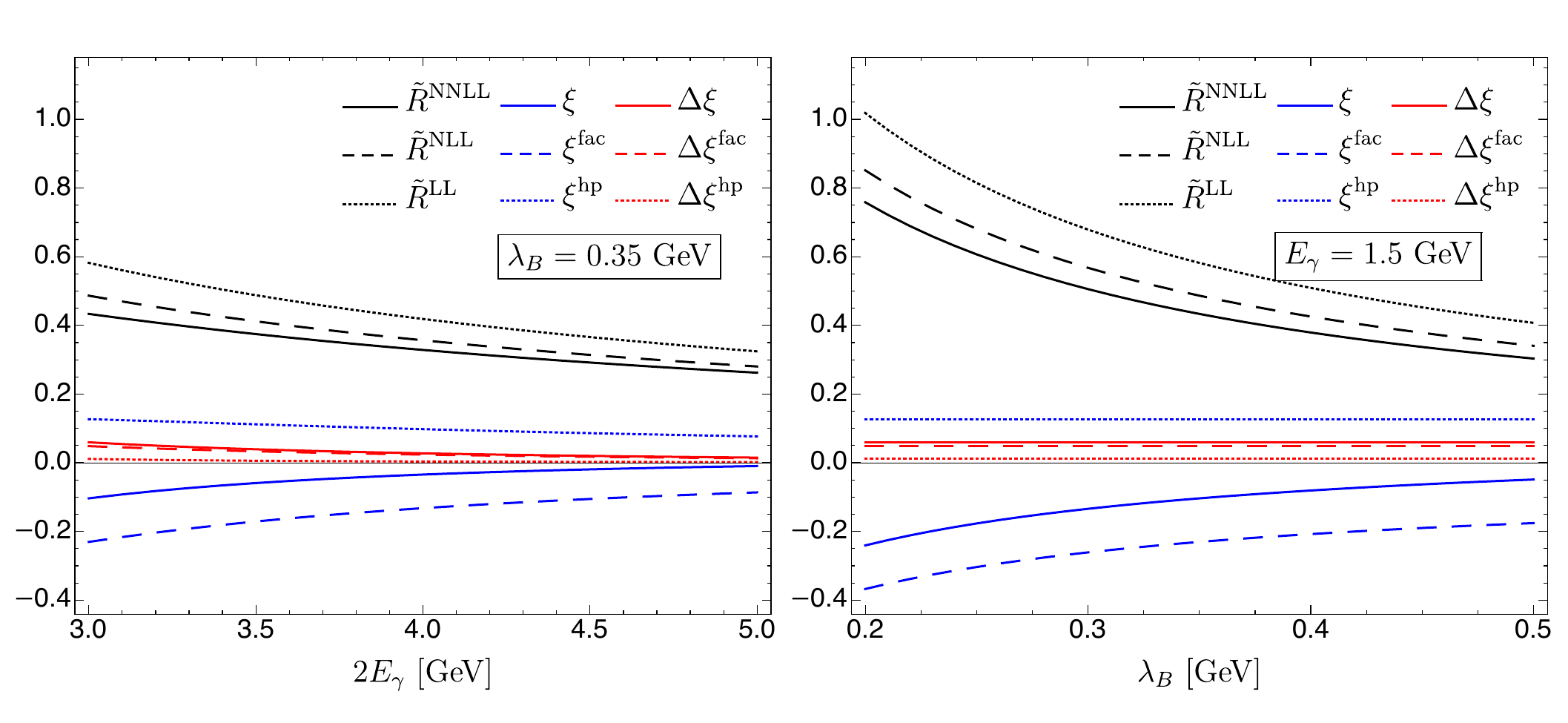}
\caption{The $E_{\gamma}$ dependence (left plot) and the $\lambda_{B}$ dependence (right plot) of the form factors where $\tilde{R}(E_{\gamma})=Q_{u}\,m_{B}/(2E_{\gamma}) \,
R(E_{\gamma})$.
}\label{fig:LPNLP}
\end{figure}

First, we focus on the LP contribution to the form factors, which includes radiative
corrections.
Fixing the photon energy to $E_{\gamma}=2.2$~GeV,
we show the $\lambda_B$ and $\sigma^{(n)}_B$ dependence of the function $R(E_{\gamma})$
\begin{eqnarray}
R^{\rm LL}&=& \frac{f_B}{\lambda_B(\mu_s)}\,
\Big[0.859+4.15\times10^{-3}\sigma_B^{(2)}
-1.36\times10^{-4}\sigma_B^{(3)}+3.34\times10^{-6}\sigma_B^{(4)}+\cdots\Big]\,,
\nonumber \\
R^{\rm NLL}&=& \frac{f_B}{\lambda_B(\mu_s)}\,
\Big[0.696+3.03\times10^{-2}\sigma_B^{(2)}
-3.02\times10^{-3}\sigma_B^{(3)}+1.50\times10^{-4}\sigma_B^{(4)}+\cdots\Big]\,,
\nonumber \\
R^{\rm NNLL} &=& \frac{f_B}{\lambda_B(\mu_s)} \,
\Big[0.642+4.36\times10^{-2}\sigma_B^{(2)}
+3.72\times10^{-4}\sigma_B^{(3)}+3.64\times10^{-4}\sigma_B^{(4)}+\cdots\Big]\,,
\end{eqnarray}
where all the other parameters are taken their central values.
Our results are comparable to that of Ref.~\cite{Galda:2020epp}.
With $\sigma^{(n)}_B=0$, one finds that the NLL
correction reduces the LL result by about 25\%, and the NNLL result is around 10\% smaller than the NLL result.
It is also clear that the coefficients of $\sigma^{(2)}_B$ are much larger than the coefficients of $\sigma^{(3,4)}_B$ in all three cases.
With a large value for $\sigma^{(2)}_B$, this term can generate sizeable correction to the form factors.
Employing the simple exponential model of $\phi_{+}$
\begin{align}
\phi_{+}(\omega,\mu) = \frac{\omega}{\omega^{2}_{0}} \,
\exp\Big(-\frac{\omega}{\omega_{0}}\Big)\,,
\qquad
\omega_{0}=\lambda_{B}(\mu)\,,
\end{align}
one roughly obtain $\sigma^{(2)}_B\in[1.6,2.3]$.
We will utilize a relatively conserved interval $\sigma^{(2)}_B\in [0.5,2.5]$ to include also the values derived in Ref.~\cite{Galda:2020epp}.
Setting all the other parameters to their central values, we show the $E_{\gamma}$ and the $\lambda_{B}$ dependences of  the form factors $F_{V/A}$ with the LL (dotted black curve), NLL (dashed black curve),
and NNLL (solid black curve) resummation accuracy in Fig.~\ref{fig:LPNLP}.
We can see that the form factors decrease with the increase of $E_{\gamma}$ or $\lambda_{B}$.
One also observes apparently the convergent tend of the radiative corrections.

In Fig.~\ref{fig:LPNLP}, we also show the numerical impacts of the NLP contributions with the factorizable and the resolved-photon corrections separated.
It is remarkable that symmetry-preserving factorizable NLP term $\xi^{\rm fac}$ (dashed blue lines) can introduce about a 50\% decrease to the LP contribution $\tilde{R}^{\rm NNLL}$ when $E_{\gamma}$ is around 1.5~GeV.
While approximately half of this decrease will be compensated by the resolved-photon mechanism so that the total symmetry-preserving correction $\xi$ (solid blue curves) reduces $\tilde{R}^{\rm NNLL}$ by around 20\% which is of the same order as the radiative correction.
This also suggests that the factorizable and resolved-photon contributions should be calculated consistently.
The symmetry-breaking $\Delta\xi$ term (solid red curves),  which mainly from the factorizable dynamics $\Delta\xi^{\rm fac}$ (dashed red lines), results in corrections to $\tilde{R}^{\rm NNLL}$ by an amount of  $\mathcal{O}(10\%)$.

\begin{figure}[!t]
\centering
\includegraphics[width=.98\textwidth]{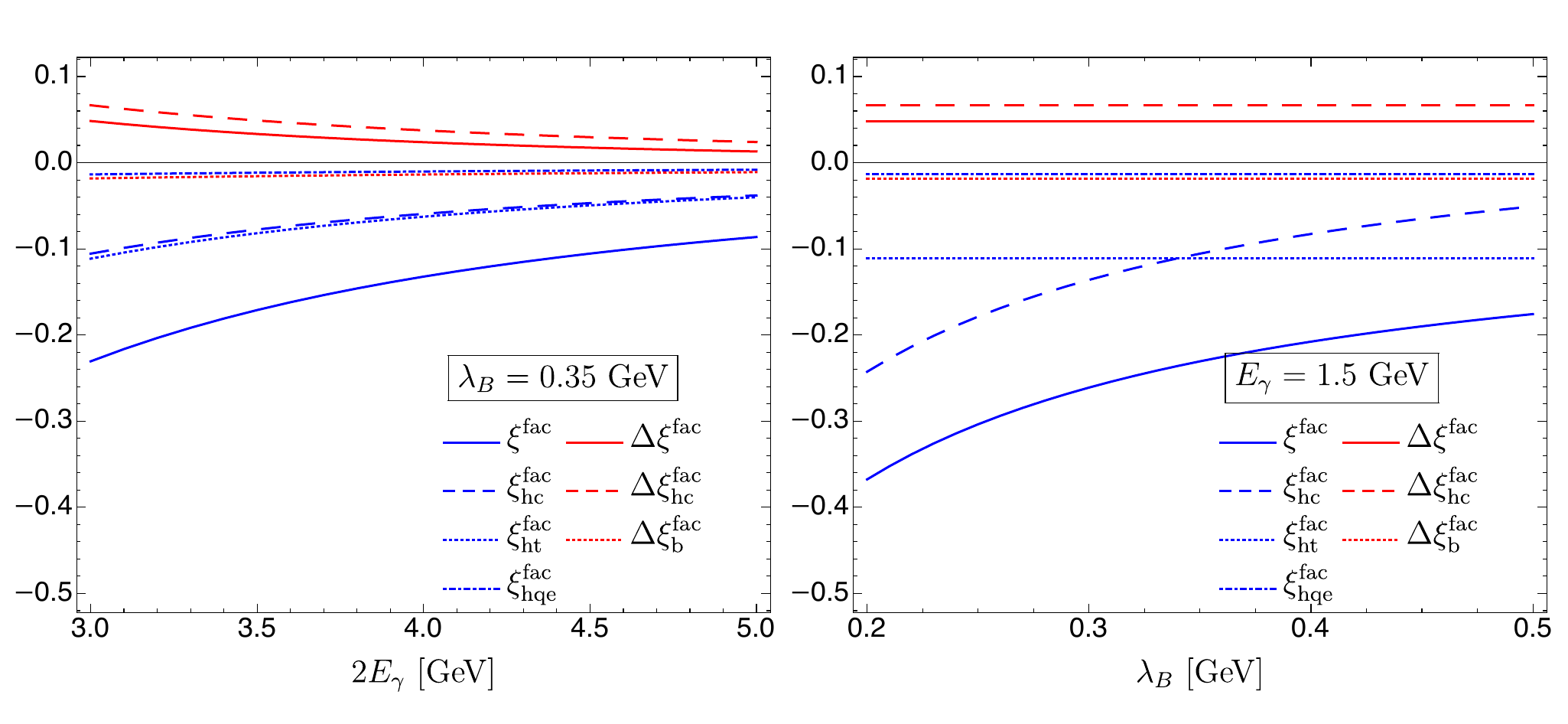}
\caption{The $E_{\gamma}$ dependence (left plot) and the $\lambda_{B}$ dependence (right plot) of the NLP factorizable form factors $\xi^{\rm fac}$ and $\Delta\xi^{\rm fac}$.}
\label{fig:fac}
\end{figure}

We illustrate in detail the effects of various factorizable NLP mechanisms in Fig.~\ref{fig:fac}.
The symmetry-preserving hard-collinear quark expansion $\xi^{\rm fac}_{\rm hc}$ (dashed blue curves) and higher-twist LCDS $\xi^{\rm fac}_{\rm ht}$ (dotted blue curves) patterns are the most significant ones, which can bring about 25\% and $15\%\sim 30\%$ reductions to the LP form factors, respectively.
The symmetry-breaking contributions are dominated by the hard-collinear quark expansion term $\Delta\xi^{\rm fac}_{\rm hc}$ (dashed red curves), which introduces about a 15\% correction to $\tilde{R}^{\rm NNLL}$.
It is surprising that the hard-collinear quark expansion dynamics alone can decrease the form factor $F_{A}$ at LP by about 40\%.
The effects of heavy-quark-expansion correction $\xi^{\rm fac}_{\rm hqe}$ (dot-dashed blue curves) and the purely symmetry-breaking correction $\Delta\xi^{\rm fac}_{\rm b}$ (dotted red curves) are tiny.

\begin{figure}[!t]
  \centering
  \includegraphics[width=.58\textwidth]{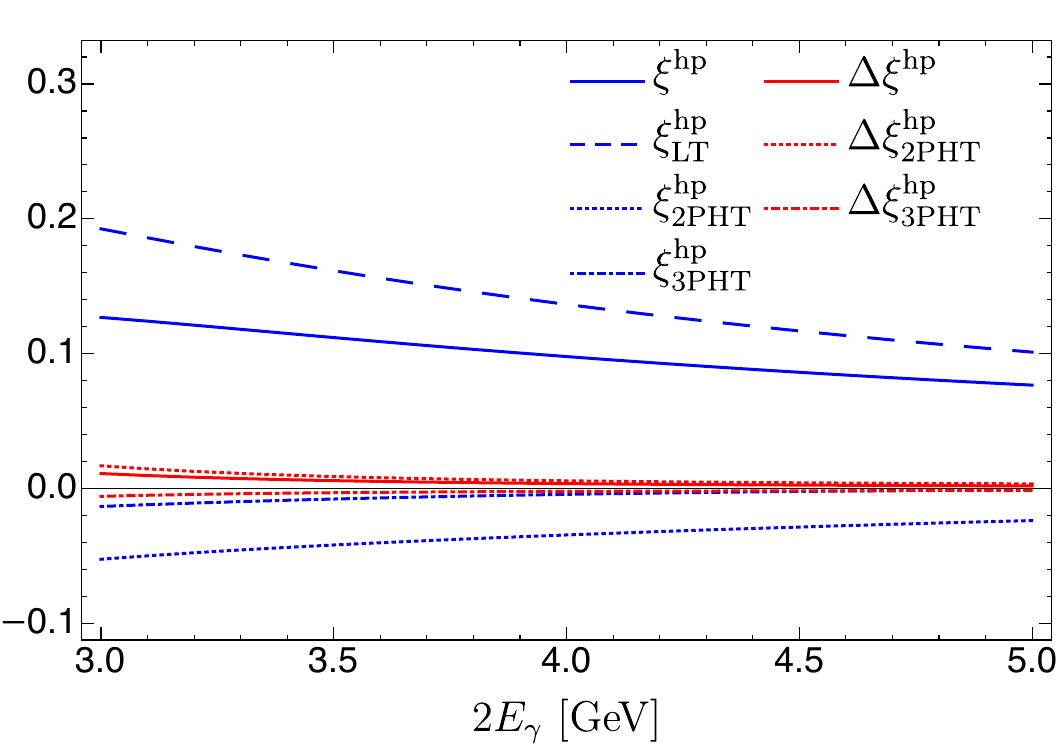}
  \caption{The $E_{\gamma}$ dependence of the resolved-photon form factors $\xi^{\rm hp}$ and $\Delta\xi^{\rm hp}$. }\label{fig:photon}
\end{figure}

The consequences of the hadronic photon contributions to the form factors are shown in Fig.~\ref{fig:photon}.
The effects of the symmetry-breaking mechanisms (red curves) are negligible.
Among the symmetry-preserving terms, only the ones corresponding to two-particle photon LCDAs are numerically significant.
It is found that the leading-twist hadronic photon contribution $\xi^{\rm hp}_{\rm LT}$ (dashed blue curves) introduces about a 50\% enhancement to the LP form factor.
Meanwhile, the two-particle higher-twist photon LCDA contribution (dotted blue curves) can decrease the LP
form factors by an amount of $\mathcal{O}(10\%)$.

\begin{figure}[!h]
\centering
\includegraphics[width=.98\textwidth]{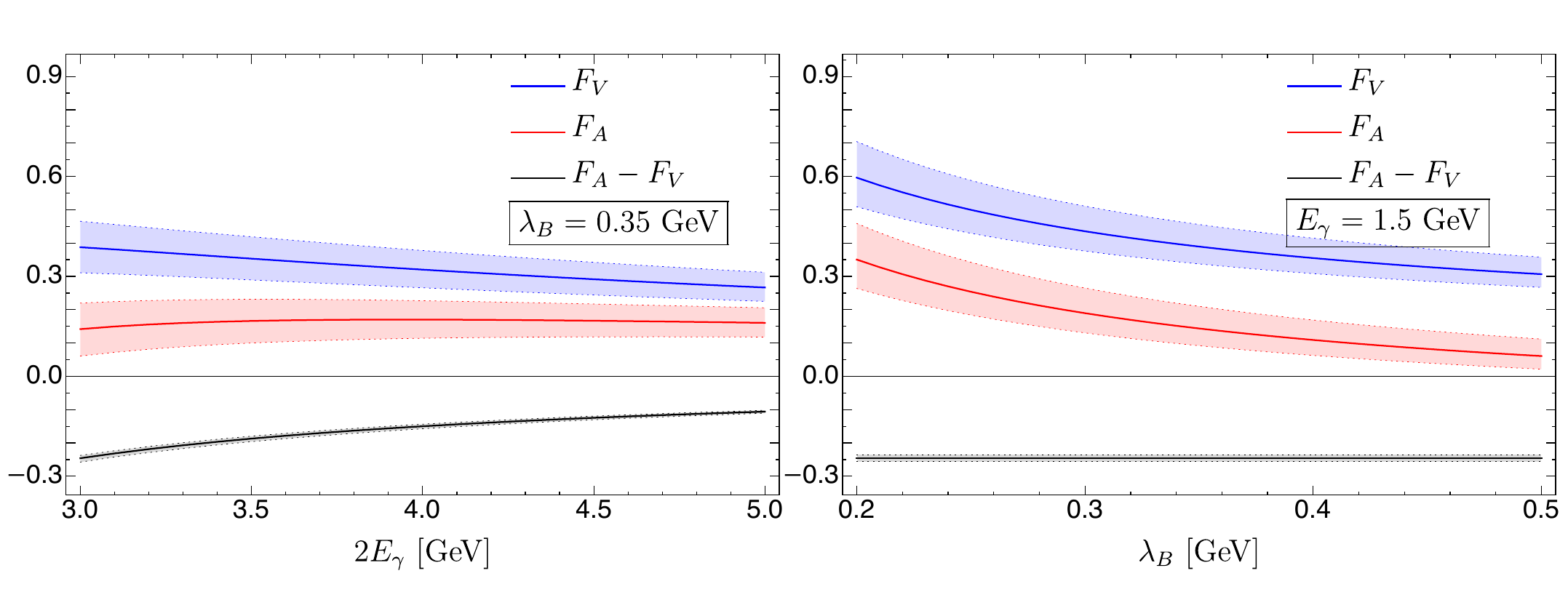}
\caption{The $E_{\gamma}$ dependence (left plot) and the $\lambda_{B}$ dependence (right plot) of the form factors $F_{V}$ and $F_{A}$.}\label{fig:FFs}
\end{figure}

\begin{table}[t]
\centering
\setlength\tabcolsep{6pt} \def\arraystretch{1.8}
\begin{tabular}{|c|c|c|c|c|c|c|c|c|c|c|c|}
\hline
~ &  Total & $\mu$ & $\mu_{h1}$ & $\sigma_B^{(2)}$
& $2\lambda_E^2+\lambda_H^2$ & $\widehat{\omega}$ & $\langle\bar{q}q\rangle$ & $s_0$ & $M^2$ \\
  \hline
  $F_V$ &  $0.387^{+0.078}_{-0.077}$ & $^{+0.000}_{-0.021}$ & $^{+0.047}_{-0.028}$ & $^{+0.029}_{-0.029}$
  &  $^{+0.021}_{-0.017}$ & $^{+0.020}_{-0.000}$ & $^{+0.036}_{-0.031}$
   & $^{+0.011}_{-0.013}$ & $^{+0.013}_{-0.019}$ \\
  \hline
  $F_A$ &  $0.142^{+0.078}_{-0.081}$  & $^{+0.000}_{-0.021}$ & $^{+0.047}_{-0.028}$ & $^{+0.029}_{-0.029}$
  & $^{+0.021}_{-0.017}$ & $^{+0.020}_{-0.000}$ & $^{+0.030}_{-0.026}$
   & $^{+0.010}_{-0.013}$ & $^{+0.017}_{-0.027}$ \\
  \hline
\end{tabular}
\caption{Theoretical predictions for the form factors with $E_\gamma=1.5$~GeV and $\lambda_B$ fixed at $0.35$~GeV.
The total uncertainties, obtained by adding all separate uncertainties in quadrature, as well as the numerically important individual uncertainties are displayed.}
\label{tab:err}
\end{table}

\begin{figure}[!h]
\centering
%Requires \usepackage{graphicx}
\includegraphics[width=.58\textwidth]{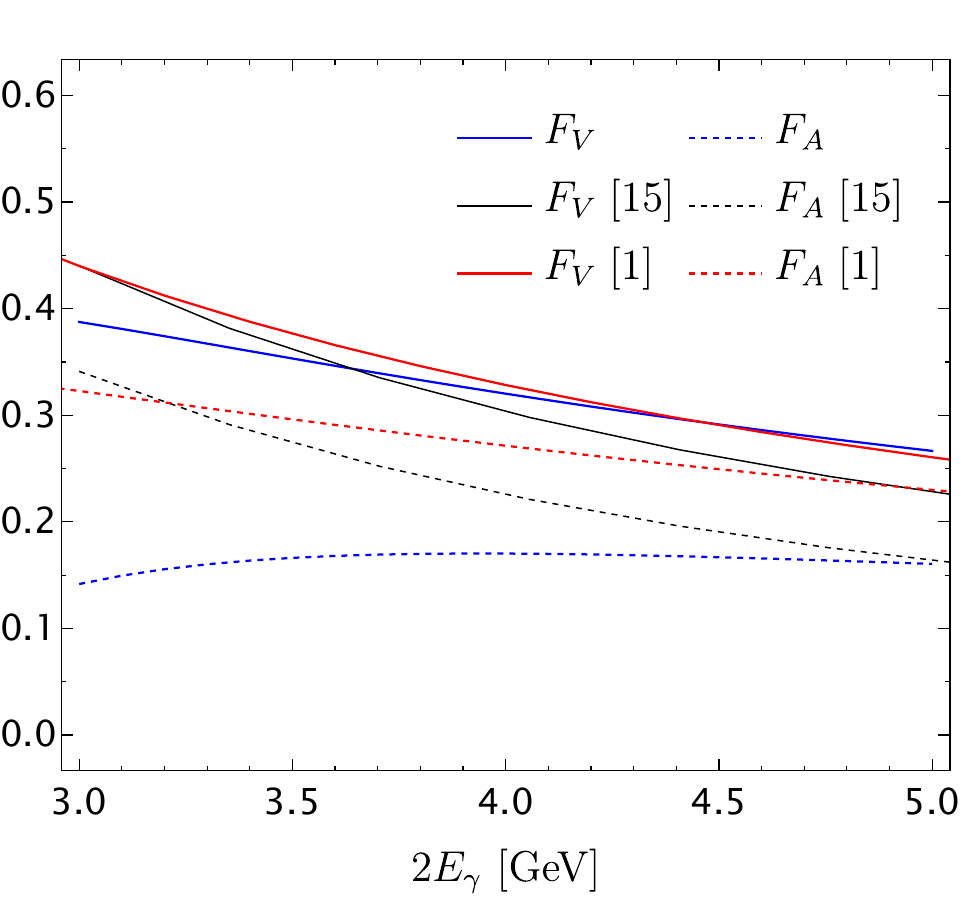}
\caption{The $E_{\gamma}$ dependence of the form factors $F_V$ (solid line) and $F_A$ (dashed line) from this work , Ref.~\cite{Janowski:2021yvz} and Ref.~\cite{Beneke:2018wjp} represented by blue, black and red line respectively.
}\label{fig:FFvs}
\end{figure}

We show our final predictions for the form factors $F_{A}$ (red curves) and $F_{V}$ (blue curves) as well as their
difference (black curves) in Fig.~\ref{fig:FFs}. The theoretical uncertainties, which are obtained by adding in quadrature the separate uncertainties caused by individual variations of all input parameters, are indicated by the bands.
The difference between the form factors $F_{A}-F_{V}$ is always negative because it is dominant by two terms $-2\Delta\xi^{\rm fac}_{\rm hc}$ and $Q_{\ell}f_{B}/E_{\gamma}$, as one can see from the above analysis.
It decreases fast with $E_{\gamma}$ decreases in the small-$E_{\gamma}$ region since $\Delta\xi^{\rm fac}_{\rm hc}$ is proportional to $1/E^{2}_{\gamma}$.
In the region where $\lambda_{B}$ is large and $E_{\gamma}$ is small, this difference will compensate the positive form factor $F_{V}$ so that $F_{A}$ can be negligible, see the right plot.
Fixing $E_\gamma=1.5$~GeV and $\lambda_B=0.35$~GeV, we also show our predictions for the form factors with uncertainties in Tab.~\ref{tab:err}.
%{\color{red} 
We also show in Fig.~\ref{fig:FFvs} the form factors $F_V$ (solid line) and $F_A$ (dashed line) from this work , Ref.~\cite{Janowski:2021yvz} and Ref.~\cite{Beneke:2018wjp} (with $\hat{\sigma}_1=0$ and 
$\lambda_B=0.35$) represented by blue, black and red line respectively. The form factors from Ref.~\cite{Janowski:2021yvz} can be reproduced by using BCL parameterization and associated fit parameters. Interestingly, the form factor $F_V$ from Ref.~\cite{Beneke:2018wjp,Janowski:2021yvz} are a little bit smaller than that of our predicton at large photon energy region, in contrast, the form factor $F_A$ from Ref.~\cite{Beneke:2018wjp,Janowski:2021yvz} is much greater than our prediton. This interesting pattern can be partly attributed to the very fact that the newly derived hard-collinear quark expansion contribution generate the most of symmetry-breaking contributions.
%}

\subsection{Phenomenological applications}

\begin{figure}[!h]
\centering
%Requires \usepackage{graphicx}
\includegraphics[width=.98\textwidth]{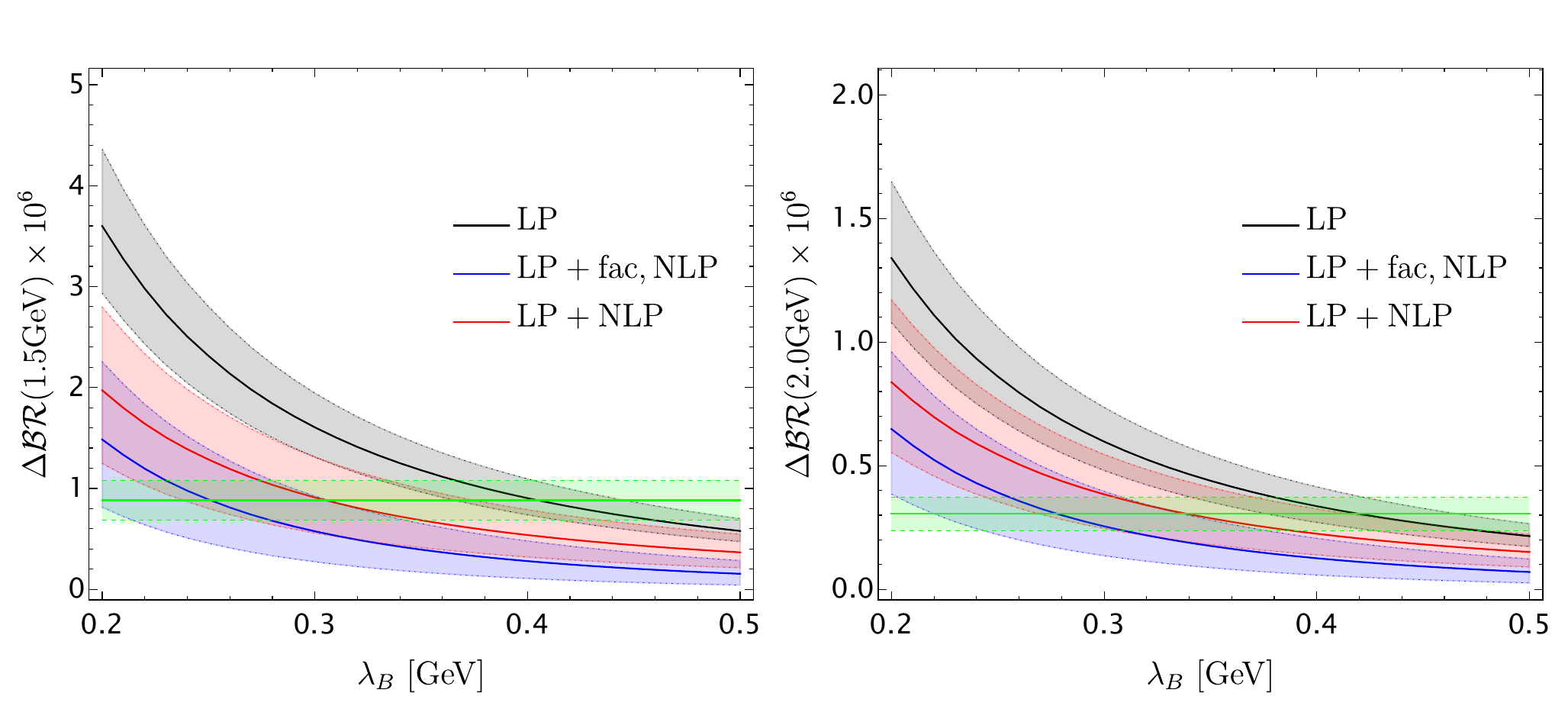}
\caption{The $\lambda_B$ dependence of partial branching fractions $\Delta\mathcal{BR}(1.5~{\rm GeV})$ (left plot) and  $\Delta\mathcal{BR}(2.0~{\rm GeV})$ (right plot).
}\label{fig:BR}
\end{figure}

Armed with our predictions for the $\decay$
form factors, we proceed to discuss the constraints on the inverse moment $\lambda_B$
taking advantage of the future measurements of the partial branching fractions with a photon-energy cut
to get rid of the soft photon radiation.
With the help of the differential decay width \eqref{eq:br},
one derives the partial branching fraction with a phase-space cut on the
photon energy
\begin{eqnarray}
\Delta{\cal BR}(E_{\rm cut})
= \tau_B \, \int_{E_{\rm cut}}^{m_B/2} \, d E_{\gamma} \,
{d \, \Gamma \over d \, E_{\gamma}} \,,
\end{eqnarray}
where $\tau_B$ is the lifetime of the $B$ meson.

The newly improved observation from the Belle collaboration shows that the partial branching fraction for
$B^+\to\gamma\ell^+\nu_{\ell}$ with $E_{\gamma}>1$~GeV is
$\Delta{\cal BR}(1~{\rm GeV})=(1.4\pm1.0\pm0.4)\times 10^{-6}$
and the Bayesian upper limit is
$\Delta{\cal BR}(1~{\rm GeV})<3.0\times 10^{-6}$
at 90\% confidence level~\cite{Gelb:2018end}.
Because that the $1/E_{\gamma}$ power expansion becomes not
reliable in the small-$E_{\gamma}$ region~\cite{Beneke:2018wjp},
experimental data with a larger photon energy cut, like 1.5 or 2.0~GeV,
is needed for a precise determination of the inverse moment.
We show the $\lambda_B$ dependence of the partial branching fractions of
the radiative leptonic decay in Fig.~\ref{fig:BR} for $E_{\rm cut}>1.5$~GeV and $E_{\rm cut}>2$~GeV.
It can be observed that
the partial branching fraction grows  with
the decrease of $\lambda_B$, which is mainly due to the reciprocal dependence of
the LP form factors on the inverse moment.
We also find that the NLP effects (red curves) generate about $40\%\sim 50\%$ decreases to the partial branching fractions at LP (black curves).
%{\color{red}  
We also show (green band) in Fig.~\ref{fig:BR} the $\Delta \mathcal{BR}$(1.5 GeV) and $\Delta \mathcal{BR}$(2.0 GeV) from table 7 of Ref~\cite{Janowski:2021yvz}, it is interesting to see that the newly derived hard-collinear quark expansion contribution lower down the $F_A$, may extract larger value of inverse moment of B-meson LCDAs. We notice that the uncertainty of the NLP prediction is not reduced compared to the LP one. 
This is because the uncertainty from the NLP effects is not included in the LP curves.
%}

\begin{figure}[!t]
\centering
\includegraphics[width=.98\textwidth]{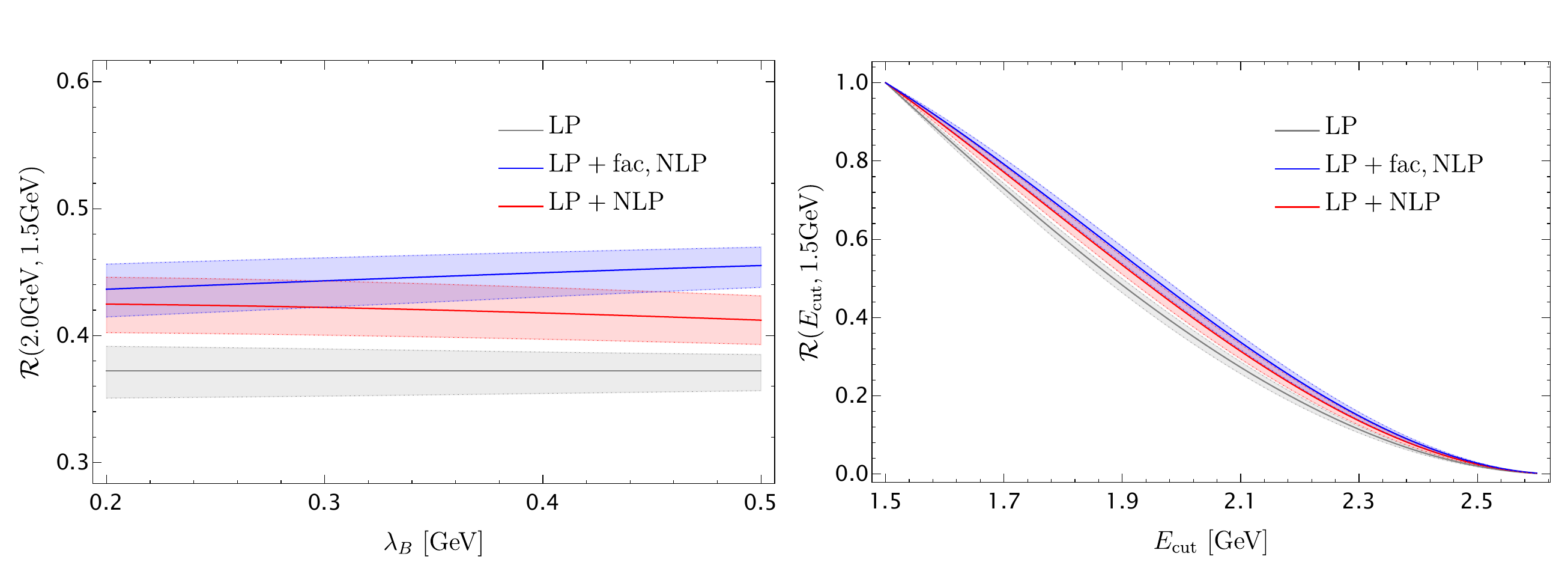}
\caption{The $\lambda_{B}$ dependence of
$\mathcal{R}(2.0~{\rm GeV},1.5~{\rm GeV})$ and the $E_{\rm cut}$ dependence of $\mathcal{R}(E_{\rm cut},1.5~{\rm GeV})$.}
\label{fig:R}
\end{figure}

To eliminate the theoretical uncertainties originating from our
prediction, we define the ratio of partial branching fractions
\begin{equation}\label{R-BR}
\mathcal{R}(E_{\gamma 1},E_{\gamma 2})=
\frac{\Delta{\cal BR}(E_{\gamma 1})}
{\Delta{\cal BR}(E_{\gamma 2})}\,.
\end{equation}
We show our predictions for this ratio in Fig.~\ref{fig:R}.
One can conclude from the left plot that the uncertainty of $\mathcal{R}(2.0~{\rm GeV},1.5~{\rm GeV})$ is of $\mathcal{O}(5\%)$ and also the ratio is almost independent of $\lambda_B$.
The effect of the NLP corrections to the $\mathcal{R}$ is significant, it introduces about a 10\% increase to the ratio at LP.
In the right plot, we also illustrate the $E_{\rm cut}$ dependence of $\mathcal{R}(E_{\rm cut},1.5~{\rm GeV})$, which equals 1 at $E_{\rm cut}=1.5$~GeV and 0 at $E_{\rm cut}=m_B/2$.
We hope that our prediction could be checked by future experiments.

\section{Conclusion}\label{sec:Conclusion}

In this paper, we reconsidered the $\decay$ decay for photon with large energy $E_\gamma \sim \mathcal{O}(m_b)$.
First, we formally re-derived the NNLL-resummation improved factorization formula for the form factors at LP in the Laplace space.
In the derivation, we chose the factorization scale to be a hard-collinear scale such that the jet function is free of large logarithms.
Employing this resummation formula, we performed complete NNLL resummations to the hard function and the static $B$-meson decay constant.
While because of the yet unknown three-loop anomalous dimension of the leading-twist $B$-meson LCDA, we could only fulfill a partial NNLL resummation to the LCDA.

Second, we calculated various NLP contributions to the decay form factors.
We computed the symmetry-preserving NLP correction from hard-collinear quark propagator $\xi^{\rm fac}_{\rm hc}$ at the tree level by using operator-level identities from the equation of motions of quarks.
The operator identities are essential when we meet operators, such as $[\not\!\partial\bar{q}(x)]\, \Gamma \, h_{v}(0)$ and $[v\cdot\partial\,\bar{q}(x)]\, \Gamma \, h_{v}(0)$, whose fields are not separated by light-like distances.
Applying the operator identities, we reproduced the LO contribution from the subleading term of the $b$-quark field in the heavy-quark expansion $\xi^{\rm fac}_{\rm hqe}$.
In the framework of the LCSR, we updated the resolved photon contributions $\xi^{\rm hp}$ and $\Delta\xi^{\rm hp}$, which includes leading-twist photon LCDA contribution at the one-loop level as well as LO correction from higher-twist photon LCDAs, with the corrected definitions of photon LCDAs.
For completeness, we also collected symmetry-breaking NLP corrections from the photon emission off the $b$ quark $\Delta\xi^{\rm fac}_{\rm b}$ and hard-collinear quark propagator $\Delta\xi^{\rm fac}_{\rm hc}$~\cite{Beneke:2011nf} and symmetry-preserving correction from the higher-twist $B$-meson LCDAs $\xi^{\rm fac}_{\rm ht}$~\cite{Beneke:2018wjp}.

Third, employing the above derived LP and NLP contributions, we investigated their numerical impacts on the form factors.
The corrections from the NLO contribution in $\alpha_s$ and NLP contribution in $\Lambda_{\rm QCD}/m_b$ to the form factors are rough of the same size, the former generates about $25\%$ decrease to both $F_V$ and $F_A$, and the later brings about a $10\%$ ($30\%$) decrease to the form factor $F_V$ ($F_A$).
Among all the NLP contributions, the resolved photon correction from leading twist-2 photon LCDA is numerically most important, it increases the LP form factors by around 50\%.
The correction from the hard-collinear quark expansion introduces a 30-40\% reduction to the form factor $F_A$.
The symmetry-preserving contribution from the higher-twist LCDAs decreases the form factors by an amount of $\mathcal{O}(20)\%$.
All the other mechanisms are numerically not that significant.

Last but not least, we analyzed the partial branching faction $\Delta{\cal BR}$ as well as the ratio of partial branching factions $\mathcal{R}$ of the $\decay$ decay.
For our results to be reliable, we introduced a large photon-energy cut $E_\gamma>1.5$~GeV or $E_\gamma>2.0$~GeV to the $\decay$ decay.
We found that the partial branching faction depends strongly on the inverse moment $\lambda_B$.
While the newly introduced ratio $\mathcal{R}$ is almost independent of $\lambda_B$.
The uncertainties of the partial branching factions will to a large extent cancel in the ratio, and this leads to a small, about 5\%, uncertainty to the ratio $\mathcal{R}$.

The various NLP contributions to the form factors could be re-analyzed in the framework of SCET, where different dynamics receive definite operator-level definitions.
This will deepen our knowledge of the $\decay$ decay.
The NLO corrections to the various subleading-power contributions, which could numerically be of the same size as the NNLO corrections at LP, are of great importance.
These corrections are also dispensable for a better understanding of exclusive $B$-meson decays.

\subsection*{Acknowledgements}
%Y.M.W. acknowledges support from the National Youth Thousand Talents Program, the Youth Hundred Academic Leaders Program of Nankai University, the National Natural Science Foundation of China with Grant No. 11675082 and 11735010, and the Natural Science Foundation of Tianjin with Grant No. 19JCJQJC61100.
 The research of B.Y.C. is supported by the National Natural Science Foundation of China with Grant No. 12405109 and Research Foundation of Chongqing University of Science and Technology under the project No.ckrc20231220. Y.L.S. is supported by the National Natural Science Foundation of China with Grants No. 12175218 and 12435004
and the Natural Science Foundation of Shandong with Grants No. ZR2024MA076 and ZR2022ZD26. C.W is supported in part by the National Natural Science Foundation of China with Grant No.12105112 and the Natural Science Foundation of Jiangsu Education Committee with Grant No.
21KJB140027. Y.B.W. is supported in part by the National Natural Science Foundation of China with Grant No.12305099 and the Alexander-von-Humboldt Stiftung.

\appendix
\section{$B$-meson LCDA}
\label{DAofB}

The definitions of the two-particle and three-particle $B$-meson LCDAs are~\cite{Braun:2017liq}
\begin{align}
& \langle \,0\, |\, \bar{q}_{\alpha} (x) \,[x,0]\,
h_{v\beta}(0)\,|\, \bar B(v)\,\rangle
\nonumber \\
=& - \frac{i \tilde f_B(\mu) \, m_B}{4} \,
\bigg [  \frac{1+\!\not\! v}{2} \, \,
\bigg\{ 2 \, \Big ( \Phi_{+}(v\cdot x) + x^2 \, G_{+}(v\cdot x)  \Big )
\nonumber \\
& - {\not\! x \over v \cdot x}  \,
\Big[ \Big( \Phi_{+}(v\cdot x) - \Phi_{-}(v\cdot x)  \Big)
+  x^2 \, \Big( G_{+}(v\cdot x) - G_{-}(v\cdot x)  \Big)   \Big]   \bigg\}
\, \gamma_5\, \bigg ]_{\beta\alpha} \,, \\
\nonumber \\
&
\langle \, 0 \,|\, \bar q_{\alpha}(t_1 \, \bar{n}) \,
[t_1\bar{n},t_2\bar{n}] \, g_s \, G_{\mu \nu}(t_2 \, \bar{n}) \,
[t_2\bar{n},0]\, h_{v \, \beta}(0) \,|\, \bar B(v) \,\rangle \nonumber \\
=&~ {\tilde{f}_B(\mu) \, m_B \over 2} \,
\bigg [ \frac{1 + \!\not\! v}{2} \, \bigg \{ (v_{\mu}\,\gamma_{\nu} - v_{\nu}\,\gamma_{\mu})  \,
\Big(\Psi_A - \Psi_V \Big )
- i \, \sigma_{\mu \nu} \, \Psi_V
 - (\bar{n}_\mu \, v_{\nu} - \bar{n}_\nu \, v_{\mu} ) \, X_A
\nonumber  \\
& + (\bar{n}_\mu \, \gamma_{\nu} - \bar{n}_\nu \, \gamma_{\mu} ) \,
\Big ( W  + Y_A   \Big )  + \, i \, \epsilon_{\mu \nu \alpha \beta} \,
\bar{n}^{\alpha} \, v^{\beta}  \, \gamma_5 \, \tilde{X}_A
- \, i \, \epsilon_{\mu \nu \alpha \beta} \,
\bar{n}^{\alpha} \, \gamma^{\beta}  \, \gamma_5 \, \tilde{Y}_A
  \nonumber \\
&- \, (\bar{n}_\mu \, v_{\nu} -  \bar{n}_\nu \, v_{\mu} )
\not\!{\bar{n}} \, W
+ \, (\bar{n}_\mu \, \gamma_{\nu} - \bar{n}_\nu \, \gamma_{\mu} )
\not\! \bar{n} \, Z   \bigg \}  \, \gamma_5 \,\bigg ]_{\beta \, \alpha}(t_1, t_2)  \,,
\label{eq:LCDA_def}
\end{align}
where $G_{\mu\nu}=i/g_s[D_\mu,D_\nu]$ (with $D_{\mu} = \partial_\mu - i g_s A_\mu$).
The momentum-space two-particle and three-particle LCDAs are related to the position-space LCDAs by
\begin{align}
\Phi_{\rm 2P}(t)=&
\int_0^\infty d\omega\,e^{-i\omega t}\,\phi_{\rm 2P}(\omega) \,,
\nonumber\\
\Phi_{\rm 3P}(t_1,t_2)=&
\int_0^\infty d\omega_1\,d\omega_2\,
e^{-i\omega_1 t_1-i\omega_2t_2}\,\phi_{\rm 3P}(\omega_1,\omega_2) \,,
\end{align}
where uppercase letters and lowercase letters represent position-space LCDAs and momentum-space LCDAs, respectively.
The three-particle LCDAs can be expressed by the LCDAs with definite twists
\begin{align}
\Phi_{3} = &~ \Psi_{A} - \Psi_{V}\,,
& \Psi_{A} = &~ \frac{1}{2}\, (\Phi_{3}+\Phi_{4})\,,
\nonumber
\\
\Phi_{4} = &~ \Psi_{A} + \Psi_{V}\,,
& \Psi_{V} = &~ \frac{1}{2}\, (-\Phi_{3}+\Phi_{4})\,,
\nonumber
\\
\Psi_{4} = &~ \Psi_{A} + X_{A}\,,
& X_{A} = &~ \frac{1}{2}\, (-\Phi_{3}-\Phi_{4}+2\,\Psi_{4})\,,
\nonumber
\\
\tilde{\Psi}_{4} = &~ \Psi_{V} - \tilde{X}_{A}\,,
& \tilde{X}_{A} = &~ \frac{1}{2}\, (-\Phi_{3}+\Phi_{4}-2\,\tilde{\Psi}_{4})\,,
\nonumber
\\
\Phi_{5} = &~ \Psi_{A} + \Psi_{V} + 2\,Y_{A} -2\,\tilde{Y}_{A} +2\,W\,,
& Y_{A} = &~ \frac{1}{2}\, (-\Phi_{3}-\Phi_{4}+\Psi_{4}-\Psi_{5})\,,
\nonumber
\\
\Psi_{5} = & - \Psi_{A} + X_{A} -2\,Y_{A}\,,
& \tilde{Y}_{A} = &~ \frac{1}{2}\, (-\Phi_{3}+\Phi_{4}
-\tilde{\Psi}_{4}+\tilde{\Psi}_{5})\,,
\nonumber
\\
\tilde{\Psi}_{5} = & - \Psi_{V} - \tilde{X}_{A} + 2\,\tilde{Y}_{A}\,,
& W = &~ \frac{1}{2}\, (\Phi_{4}-\Psi_{4}-\tilde{\Psi}_{4}
+\Phi_{5}+\Psi_{5}+\tilde{\Psi}_{5})\,,
\nonumber
\\
\Phi_{6} = &~ \Psi_{A} - \Psi_{V} + 2\,Y_{A} + 2\,\tilde{Y}_{A} +2\,W - 4\,Z\,,
& Z = &~ \frac{1}{4}\, (-\Phi_{3}+\Phi_{4}-2\,\tilde{\Psi}_{4}
+\Phi_{5}+2\,\tilde{\Psi}_{5}-\Phi_{6})\,.
\end{align}
For the higher-twist LCDA $G_+$, we also separate the generic two-particle part $\hat{G}_+$ from the three-particle part
\begin{eqnarray}\label{LCDA:g-hat}
G_+(t)=\hat{G}_+(t)-\frac{1}{2}\int_0^1du\,\bar{u}\,\Psi_4(t,ut)\,.
\end{eqnarray}

\section{NNLL resummation}
\label{App:AD}

\subsection{Perturbative corrections up to NNLO}

The functions $C(E_\gamma,\mu)$, $J(E_\gamma\omega,\mu)$ and $K(\mu)$ can be calculated perturbatively.
The hard function is expanded in $\alpha_s$ as
\begin{align}
C(E_\gamma,\mu)=
\sum_{n=0}  C^{(n)}(\mu)\, \Big(\frac{\alpha_s(\mu)}{4\pi}\Big)^n\,,
\label{eq:exp.hard}
\end{align}
and the coefficients up to NNLO accuracy are~\cite{Bonciani:2008wf, Asatrian:2008uk,Beneke:2008ei,Bell:2008ws}
\begin{eqnarray}
C^{(0)}(\mu)&=&1\,,
\hspace{2cm}
C^{(1)}(\mu)=C_F\Big[ -\frac{L^2}{2}+L\,f_{-1}(u)+f_0(u)\Big]
\Big|_{u= 2E_\gamma/m_b}\,,
\nonumber\\
C^{(2)}(\mu)&=& \bigg\{F_{1,0}^{(2)}(u)
+ C_F^2 \Big[-\frac{5}{24}L^4+\frac{5}{6}L^3f_{-1}(u)+\frac{L^2}{2} \big(3f_0(u)-f_{-1}(u)\big)+L\big(2f_1(u)
\nonumber\\
&-&f_{-1}(u)f_0(u)\big)
-f_{-1}(u)f_1(u)+f_2(u)\Big]
\nonumber\\
&+&C_F\bigg[ \frac{17}{18}L^3-\frac{25}{6}L^2f_{-1}(u)-\frac{L}{9}(\pi^2+87f_0(u))
-\frac{31}{3}f_1(u)+\frac{2}{9}\zeta_3
\nonumber\\
&+&n_f\Big( -\frac{L^3}{9}+\frac{L^2}{3}f_{-1}(u)+
\frac{2}{3}Lf_0(u)+\frac{2}{3}f_1(u) \Big)\bigg]
\bigg\}\Big|_{u= 2E_\gamma/m_b},
\end{eqnarray}
with $L \equiv \ln(\mu^2/m_b^2)$ and $n_f$ the number of light fermion flavours.
The function $F_{1,0}^{(2)}(u)$
is
\begin{align}
F_{1,0}^{(2)}(u)=&~ C_F^2 \Big[\frac{L^4}{3}+\frac{4}{3} L^3 \, j_{-3}(u) +2 L^2 \, j_{-2}(u)+2 L \,
	    j_{-1}(u)+j_{0}(u)\Big]
\nonumber\\
        &+3C_F\Big[-\frac{11}{9} L^3+2 L^2 \, k_{-2}(u) +2 L \, k_{-1}(u)+ k_{0}(u)\Big]
\nonumber\\
        &+C_F\frac{n_f}{2}\Big[ \frac{4}{9} L^3+2 L^2 \, p_{-2}(u) +2 L \, p_{-1}(u)+ p_{0}(u)\Big]
\nonumber\\
        &+C_F\frac{1}{2}\Big[\frac{16}{9} L^3+2 L^2 \, q_{-2}(u) +2 L \, q_{-1}(u)+ q_{0}(u)\Big]\,,
\end{align}
where the coefficient functions $f_{i}(u)$, $j_i(u)$, $k_i(u)$, $p_i(u)$ and $q_i(u)$ can be found in Ref.~\cite{Beneke:2008ei}.
The jet function and the function $K$ could be expanded similarly.
The corresponding coefficients up to NNLO level for the jet function are~\cite{Liu:2020ydl}
\begin{align}
J^{(0)}(\mu)=&~1\,,
\hspace{2cm}
J^{(1)}(\mu)=C_F \, \Big( L_p^2 - \frac{\pi^2}{6}-1 \Big)\,,
\nonumber\\
J^{(2)}(\mu)=&~C_F \,\bigg[ \frac{2}{3}\,L_p^4+\frac{2n_f-33}{9}\,L_p^3-\frac{10n_f+11\pi^2-189}{9}\,L_p^2
\nonumber\\
&+\frac{38n_f+1278\zeta_3+36\pi^2-915}{27}\,L_p
+n_f\,\Big(\frac{4\zeta_3}{9}+\frac{5\pi^2}{54}
+\frac{7}{81}\Big)
\nonumber\\
& -\frac{16\zeta_3}{3} -\frac{149\pi^4}{540}-\frac{119\pi^2}{36}+\frac{601}{54}\bigg]\,, \label{jet-expression}
\end{align}
with $L_p=\ln(2E_\gamma\,\omega/\mu^2)$,
and for the function $K$ are~\cite{Broadhurst:1994se}
\begin{eqnarray}
K^{(0)}(\mu)&=&1\,,
\hspace{2cm}
K^{(1)}(\mu)=-C_F\,\Big(\frac{3}{2}\ln\frac{\mu^2}{m_b^2}+2\Big)\,,
\nonumber\\
K^{(2)}(\mu)&=& -C_F\, \Big[\frac{9}{8}\,C_F\,\ln^2\frac{\mu^2}{m_b^2}
+\Big(3\,C_F+\frac{127}{12}+\frac{7\pi^2}{9}+\frac{5}{6}n_f\Big)
\ln\frac{\mu^2}{m_b^2}
\nonumber\\
&+&\frac{683}{48}+\frac{17\pi^2}{6}+
\frac{11\zeta_3}{3}+\frac{2\ln2}{3}\pi^2\Big]\,.
\end{eqnarray}
The expansion coefficients for the function $K^{-1}(\mu)$, which are actually needed in the factorization formula, can be obtained from $K^{(i)}$ straightforwardly.

\subsection{Anomalous dimensions}
The evolution factors $U_1(E_\gamma,\mu_{h1},\mu)$ and $U_2(\mu_{h2},\mu)$ satisfy the same RG-equation as $C(E_\gamma,\mu)$ and $\tilde{f}_B(\mu)$, respectively,
\begin{align}
\frac{d}{d\ln\mu}\,U_1(E_\gamma,\mu_{h1},\mu)
=&~ \Big[\Gamma_{\rm cusp}(\alpha_s)\,\ln\frac{2\,E_\gamma}{\mu}
+\gamma_H(\alpha_s)\Big]\,U_1(E_\gamma,\mu_{h1},\mu)\,,
\nonumber\\
\frac{d}{d\ln\mu}\,U_2(\mu_{h2},\mu)
=&~ \gamma_{hl}(\alpha_s)\,U_2(\mu_{h2},\mu)\,,
\end{align}
with initial condition $U_1(E_\gamma,\mu_{h1},\mu_{h1})=1$ and $U_2(\mu_{h2},\mu_{h2})=1$.
The anomalous dimensions and the QCD beta function are expanded as
\begin{align}
&~ \gamma_H(\alpha_s)=
\sum_{n=0}\gamma_{n,H}\,
\Big(\frac{\alpha_s}{4\pi}\Big)^{n+1}\,,
&&~
\beta(\alpha_s)=
\mu\frac{d\alpha_s}{d\mu}=
-2\alpha_s\sum_{n=0}\beta_n\,
\Big(\frac{\alpha_s}{4\pi}\Big)^{n+1}\,.
\end{align}
The other anomalous dimensions are expanded similarly to $\gamma_H$.

We collect the beta function~\cite{vanRitbergen:1997va}
\begin{align}
\beta_0=&~ 11-\frac{2}{3}n_f,\qquad
\beta_1 = 102 -\frac{38}{3}n_f, \qquad
\beta_2 = \frac{2857}{2} - \frac{5033}{18} n_f + \frac{325}{54} n_f^2,
\nonumber \\
\beta_3=&~ \frac{149753}{6} + 3564 \zeta_3
-\Big( \frac{1078361}{162} + \frac{6508}{27}\,\zeta_3 \Big)n_f
+\Big( \frac{50065}{162} + \frac{6472}{81}\,\zeta_3 \Big)n^2_f
+\frac{1093}{729} n^3_f \,,
\end{align}
the cusp anomalous dimension~\cite{Henn:2019swt}
\begin{align}
\Gamma_0 =&~4C_F\,,\quad
\Gamma_1=C_F\Big[\frac{268}{3}-4\pi^2-\frac{40}{9}n_f\Big]\,,
\nonumber
\\
\Gamma_2 =&~C_F\big[1470-\frac{536\pi^2}{3}+\frac{44\pi^4}{5}+264\zeta_3 +n_f\Big(-\frac{1276}{9}+\frac{80\pi^2}{9}-\frac{208}{3}\zeta_3\Big)
-\frac{16}{27}n_f^2\big]\,,
\nonumber
\\
\Gamma_3 =&~C_F\big[\frac{84278}{3}-\frac{44200\pi^2}{9}+
\frac{2706\pi^4}{5}-\frac{2504\pi^6}{105}+20944\zeta_3
-10824\zeta_5 -528\pi^2\zeta_3-432\zeta_3^2
\nonumber
\\
&+ n_f\big(-\frac{344345}{81}+\frac{12800\pi^2}{27}
-\frac{616\pi^4}{45}-\frac{153920\zeta_3}{27}
+\frac{19504\zeta_5}{9}+\frac{416\pi^2\zeta_3}{3}\big)
\nonumber
\\
&+n_f^2\big(\frac{17875}{243}-\frac{304\pi^2}{81}-\frac{104\pi^4}{135} +\frac{4160\zeta_3}{27}\big)+n_f^3\big(-\frac{32}{81}+\frac{64\zeta_3}{27}\Big)\big]
\nonumber
\\
&-\frac{160\pi^2}{3}-\frac{496\pi^6}{189}+\frac{320\zeta_3}{3}
-960\zeta_3^2+\frac{8800\zeta_5}{3} +\frac{n_f}{27}
\big(160\pi^2-320\zeta_3-1600\zeta_5\big)
 \,,
\end{align}
where we have used $\Gamma_n \equiv \Gamma_{n,\rm cusp}$,
the anomalous dimension of the hard function~\cite{Bruser:2019yjk}
\begin{align}
\gamma_{0,H}=&-5C_F\,,\quad \gamma_{1,H}=C_F\big[-\frac{1585}{18}
-\frac{5\pi^2}{6}+34\zeta_3+n_f\,\big(\frac{125}{27}
+\frac{\pi^2}{3}\big)\big]\,,
\nonumber
\\
\gamma_{2,H}=&~C_F\big[-\frac{307753}{324}+\frac{2941\pi^2}{54}
-\frac{2221\pi^4}{270} +\frac{13858\zeta_3}{9} -\frac{2860\zeta_5}{3}
-\frac{1544\pi^2\zeta_3}{27}
\nonumber
\\
&+ n_f\,\big(\frac{19805}{243}+\frac{781\pi^2}{81}+
\frac{17\pi^4}{405} -\frac{212\zeta_3}{27}\big) +n_f^2\,
\big(\frac{2633}{729}-\frac{10\pi^2}{27}
-\frac{8\zeta_3}{27}\big)\big]\,,
\end{align}
the anomalous dimension of HQET decay constant $\tilde{f}_B(\mu)$~\cite{Chetyrkin:2003vi}
\begin{align}
\gamma_{0,hl} =&~ 3C_F\,,\quad
\gamma_{1,hl}=-C_F\big[-\frac{127}{6}-\frac{14\pi^2}{9}+\frac{5}{3}n_f\big]\,,
\nonumber
\\
\gamma_{2,hl}=&-C_F\big[\frac{61}{4}-\frac{686\pi^2}{27}
-\frac{380\pi^4}{81} +\frac{178\zeta_3}{3}
+n_f\Big(\frac{172}{27}+\frac{196\pi^2}{81}+
\frac{332\zeta_3}{9}\Big)+n_f^2\,\frac{35}{27}\big]\,,
\end{align}
and the two-loop anomalous dimension $\gamma$ of $B$-meson LCDA~\cite{Liu:2020ydl}
\begin{align}
\gamma_0=-2C_F\, ,\quad \gamma_1=C_F\big[\frac{206}{9}-\frac{53\pi^2}{18}
-50\zeta_3+n_f\big(-\frac{32}{27}+\frac{5\pi^2}{9}\big)\big]\,.
\end{align}

\subsection{The evolution function $U(E_h,\mu_{i},\mu)$}
The solution to the general RG equation is
\begin{align}
U(E_h,\mu_{i},\mu)=&~
\exp\Big[S(\mu_{i})
-a_{\gamma_{\rm gen}}(\mu_{i})\Big]\,
\Big(\frac{E_h}{\mu_{i}}\Big)^{-a_{\Gamma}(\mu_{i})}\,,
\end{align}
where we have introduced the following functions
\begin{align}
&~ a_{\gamma_{\rm gen}}(\mu_i)\equiv
a_{\gamma_{\rm gen}}(\mu_i,\mu)=
-\int_{\alpha_s(\mu_i)}^{\alpha_s(\mu)}d\alpha\,
\frac{\gamma_{\rm gen}(\alpha)}{\beta(\alpha)}\,,
\nonumber \\
&~ S(\mu_i)\equiv
S(\mu_i,\mu)=
-\int_{\alpha_s(\mu_i)}^{\alpha_s(\mu)}d\alpha\,
\frac{\Gamma_{\rm cusp}(\alpha)}{\beta(\alpha)}
\int_{\alpha_s(\mu_i)}^{\alpha}\frac{d\hat{\alpha}}
{\beta(\hat{\alpha})}\,,
\label{rg_fun}
\end{align}
and a similar definition for $a_\Gamma \equiv a_{\Gamma_{\rm cusp}}$.
In practice, the evolution factor $U$ should be expanded in the QCD coupling constant.
The solution to the next-to-leading logarithmic (NLL) resummation accuracy can be found in Ref.~\cite{Beneke:2011nf}.
To derive the NNLL-resummation formula, the functions $S$
\footnote{
The function $S$ starts at order $\alpha^{-1}_s$.
Since the order $\alpha^{-1}_s$ term in the natural exponential function should not be expanded, it is convenient to include this term into $S^{(0)}$, see Appendix~\ref{App:AD}.
},
$a_{\gamma_{\rm gen}}$ and $a_\Gamma$ need to be expanded in $\alpha_s(\mu_i)/(4\pi)$ up to the two-loop level, and the expansion coefficients are collected in Appendix~\ref{App:AD}.
With these coefficients at hand, it is straightforward to obtain
\begin{align}
U(E_h,\mu_{i},\mu)=&~
\exp\Big[S^{(0)} - a_{\gamma_{\rm gen}}^{(0)}\Big]\,
\Big(\frac{E_h}{\mu_{i}}\Big)^{-a_\Gamma^{(0)}}
\Big[1
+U^{(1)} \, \frac{\alpha_s(\mu_{i})}{4\pi}
+ U^{(2)} \,\Big(\frac{\alpha_s(\mu_{i})}{4\pi}\Big)^2
+ \mathcal{O}(\alpha^3_s)\Big]\,,
\label{eq:U.result}
\end{align}
where
\begin{eqnarray}
U^{(1)}&=&
-\ln\frac{E_h}{\mu_{i}}\,a_\Gamma^{(1)} + S^{(1)}
-a_{\gamma_{\rm gen}}^{(1)}\,,
\nonumber\\
U^{(2)}&=&
\frac{1}{2}\ln^2\frac{E_h}{\mu_{i}}\,\Big(a_\Gamma^{(1)}\Big)^2
+\ln\frac{E_h}{\mu_{i}}\,
\Big[-a_\Gamma^{(2)}-S^{(1)}\,a_\Gamma^{(1)}
+a_{\Gamma}^{(1)}\,a_{\gamma_{\rm gen}}^{(1)}\Big]
\nonumber\\
&+& S^{(2)} - a_{\gamma_{\rm gen}}^{(2)}
+\frac{1}{2}\,\Big(S^{(1)}\Big)^2
-S^{(1)}\, a_{\gamma_{\rm gen}}^{(1)}
+ \frac{1}{2}\,\Big(a_{\gamma_{\rm gen}}^{(1)}\Big)^2
\,.
\end{eqnarray}
The $\mu$ and $\mu_i$ dependence of the factors $S^{(i)}$, $a_\Gamma^{(i)}$ and $a_{\gamma_{\rm gen}}^{(i)}$ should be understood.

\subsection{Partial NNLL resummation for $\phi_+$}
To obtain the RG equation of the LCDA $\tilde{\phi}_+$, it is convenient to introduce a Mellin transformation \footnote{
The explicit relation between LCDAs in the momentum space and Mellin space is
\begin{align}
\varphi_+(j) = \frac{-i}{2\pi} \,\Gamma(-j) \int^{\infty}_{0} d\omega\,
\Big(\frac{\omega}{\mu e^{\gamma_E}}\Big)^j \, \phi_+(\omega) \,.
\end{align}
The relation is not necessary in this work.
} and derive the RG equation for the Mellin-space LCDA $\varphi_+$~\cite{Braun:2019zhp},
\begin{align}
&~ \Big(\frac{d}{d\ln\mu}
-\Gamma_{\rm cusp}(\alpha_s)\,\frac{\partial}{\partial j}\Big)\,
\varphi_+(j,\mu)
\nonumber \\
&= -\bigg\{
\Gamma_{\rm cusp}(\alpha_s)\,\Big[
\tilde{h}(1) -\tilde{h}(1+j)
+\psi(j+2)-\psi(2)\Big]
+\gamma_+(\alpha_s)+\gamma_{hl}(\alpha_s)+j\bigg\}\,
\varphi_+(j,\mu)\,,
\nonumber \\
&~ \tilde{h}(j) \equiv \tilde{h}(\alpha_s,j)=
\int^1_0 dx \, \frac{\bar{x}^j}{x}\, h(\alpha_s,\bar x) \,,
\end{align}
where $h(\alpha_s,x)$ needs to be expanded in $\alpha_s/(4\pi)$ in the same way as the hard function, $\psi$ is the digamma function and $\bar{x}\equiv 1-x$.
The two-loop anomalous dimension of $\gamma_{+}$ and the two-loop level $h^{(1)}$  were calculated in Ref.~\cite{Braun:2019wyx}.
Actually the explicit form of $\gamma_{+}$ is not necessary in this work, and $h^{(1)}$ is
\begin{align}
h^{(1)}(x)=\ln x\,\Big[\beta_0+2C_F\,\Big(\ln x-\frac{1+x}{x}\,\ln\bar{x}-\frac{3}{2}\Big)\Big]\,,
\end{align}
where $\beta_0$ is the one-loop QCD $\beta$-function (see Appendix~\ref{App:AD}).
Employing the trivial relation between the Mellin-space LCDA and the Laplace-space LCDA~\cite{{Shen:2020hfq}}
\begin{align}
\varphi_+(j,\mu)
=&~\frac{-i}{2\pi} \,\Gamma(-j)\,
\Big(\frac{\hat\omega}{\mu e^{\gamma_E}}\Big)^{j+1}\,(\mu e^{\gamma_E}) \,
\tilde\phi_+(-j-1,\widehat\omega,\mu) \,,
\end{align}
one obtains the RG equation of $\tilde\phi_+$ easily~\cite{Galda:2020epp}
\begin{align}
&~ \left(\frac{d}{d\ln\mu}+\Gamma_{\rm cusp}(\alpha_s)\frac{\partial}{\partial\eta} \right)
\tilde{\phi}_+(\eta,\widehat\omega,\mu) \nonumber \\
=&~ \left\{\Gamma_{\rm cusp}(\alpha_s)\left[
\ln\frac{\widehat{\omega}}{\mu}
+\tilde{h}(-\eta)
-\psi(1+\eta) -\psi(1-\eta)
-2\gamma_E \right]-
\gamma(\alpha_s)\right\}\tilde{\phi}_+(\eta,\widehat\omega,\mu)\,,
\end{align}
where
\begin{align}
\gamma(\alpha_s) = \gamma_+(\alpha_s) + \gamma_{hl}(\alpha_s)
+\Gamma_{\rm cusp}(\alpha_s)\,
\left[\tilde{h}(1)-1\right] \,.
\end{align}
The two-loop anomalous dimension of $\gamma$ is determined by that of $\gamma_{+}$ and $h^{(1)}$.
The explicit expression for $\gamma$ is collected in Appendix~\ref{App:AD}.

The explicit solution to the RG equation \eqref{eq:RG.DA} at $\alpha_s$ accuracy can be found in Ref.~\cite{Galda:2020epp}.
We would like to derive the formal $\alpha^2_s$-level solution, which could be employed to fulfill the NNLL resummation to the LCDA once the three-loop anomalous dimensions of $\gamma_{+}$ and $h$ are known.
Since any combination of $\eta+a_\Gamma(\mu_s,\mu)$ is a solution to \eqref{eq:RG.DA} with the right-hand-side terms set to zero, we separate the RG-evolution factor into $\eta$-independent $U_{\phi1}$ and $\eta$-dependent $U_{\phi2}$ parts.
And the combination
$U_{\phi1}(\widehat\omega,\mu_s,\mu)\,
U_{\phi2}(\eta,\mu_s,\mu)\,
\tilde{\phi}_+(\eta+a_\Gamma(\mu_s),\widehat\omega,\mu_s)$
will be the solution to the RG equation satisfied by $\tilde{\phi}_+$.

The factor $U_{\phi1}$ satisfies the RG equation
\begin{align}
\frac{d}{d\ln\mu}\,
U_{\phi1}(\widehat\omega,\mu_s,\mu)
=&~ \left\{
\Gamma_{\rm cusp}(\alpha_s) \, \ln\frac{\widehat{\omega}}{\mu}
- \Big[ 2\gamma_E \,\Gamma_{\rm cusp}(\alpha_s) +
\gamma(\alpha_s) \Big]
\right\}
U_{\phi1}(\widehat\omega,\mu_s,\mu) \,,
\end{align}
with initial condition $U_{\phi1}(\widehat\omega,\mu_s,\mu_s)=1$.
The equation is closely related to the general RG equation \eqref{eq:gen.RG}, and we have
\begin{align}
U_{\phi1}(\widehat\omega,\mu_s,\mu)=
U(\widehat\omega,\mu_{s},\mu)
\Big|_{\gamma_{\rm gen}\to
- (\gamma+2\gamma_E \Gamma_{\rm cusp})} \,.
\end{align}
The factor $U_{\phi2}$ is expressed as
\begin{align}
U_{\phi2}(\eta,\mu_s,\mu)
=&~
\exp\bigg[
G(\eta,\mu_s,\mu)+
\int^{a_\Gamma(\mu_s)}_0 d\hat{a}_\Gamma \,
\Big(\psi(1+\eta+\hat{a}_\Gamma)
+\psi(1-\eta-\hat{a}_\Gamma)\Big)
\bigg]
\nonumber \\
=&~
\exp\Big[
G(\eta,\mu_s,\mu)
\Big]\,R_\Gamma(\eta,\mu_s,\mu)
\,,
\nonumber \\
G(\eta,\mu_s,\mu)=&~
\int\limits_{\alpha_s(\mu_s)}^{\alpha_s(\mu)}
\frac{d\alpha}{\beta(\alpha)}\,
\Gamma_{\rm cusp}(\alpha) \,
\tilde{h}(\alpha,-\eta-a_\Gamma(\mu_\alpha))\,,
\nonumber \\
R_\Gamma(\eta,\mu_s,\mu)
=&~
\frac{\Gamma\big(1+\eta+a_\Gamma(\mu_s)\big)\,
\Gamma(1-\eta)}{\Gamma\big(1-\eta-a_\Gamma(\mu_s)\big)\,\Gamma(1+\eta)} \,,
\end{align}
with $\mu_{\alpha}$ satisfing $\alpha_{s}(\mu_{\alpha})\equiv \alpha$.

Having the evolution factors $U_{\phi1}$ and $U_{\phi2}$ at hand, the large logarithms appearing in the LCDA $\tilde{\phi}_+$ can be resummed directly.
The evolution function $\tilde{\phi}_{+,U}$ is then
\begin{align}
\tilde{\phi}_{+,U}(\eta,\mu_s,\mu)=&~
\bigg( \frac{2E_\gamma\,\hat\omega}{\mu^2} \bigg)^\eta \,
U_{\phi1}(\widehat\omega,\mu_s,\mu)\,
U_{\phi2}(-\eta,\mu_s,\mu)\,
\tilde{\phi}_+(-\eta+a_\Gamma(\mu_s),\widehat\omega,\mu_s)\,.
\end{align}

\subsection{Expansion of evolution factors}
For completeness, we recollect the definitions of functions entering the evolution factors
\begin{align}
&~  a_{\gamma_{\rm gen}}(\mu_i,\mu)=
-\int_{\alpha_s(\mu_i)}^{\alpha_s(\mu)}d\alpha\,
\frac{\gamma_{\rm gen}(\alpha)}{\beta(\alpha)}
& =&~
\sum_{n=0}  a_{\gamma_{\rm gen}}^{(n)}\,
\Big(\frac{\alpha_s(\mu_i)}{4\pi}\Big)^n\,,
\nonumber \\
&~ S(\mu_i,\mu)=
-\int_{\alpha_s(\mu_i)}^{\alpha_s(\mu)}d\alpha\,
\frac{\Gamma_{\rm cusp}(\alpha)}{\beta(\alpha)}
\int_{\alpha_s(\mu_i)}^{\alpha}\frac{d\hat{\alpha}}
{\beta(\hat{\alpha})}
& =&~
\sum_{n=0}  S^{(n)}\,
\Big(\frac{\alpha_s(\mu_i)}{4\pi}\Big)^n\,,
\nonumber \\
&~G(\eta,\mu_i,\mu)=
\int\limits_{\alpha_s(\mu_i)}^{\alpha_s(\mu)}
\frac{d\alpha}{\beta(\alpha)}\,
\Gamma_{\rm cusp}(\alpha) \,
\tilde{h}(\alpha,-\eta-a_\Gamma(\mu_\alpha))
& =&~
\sum_{n=0}  G^{(n)}\,
\Big(\frac{\alpha_s(\mu_i)}{4\pi}\Big)^n\,,
\nonumber \\
&~ R_\Gamma(\eta,\mu_i,\mu)
=
\frac{\Gamma\big(1+\eta+a_\Gamma(\mu_i)\big)\,\Gamma(1-\eta)}
{\Gamma\big(1-\eta-a_\Gamma(\mu_i)\big)\,\Gamma(1+\eta)}
& =&~
\sum_{n=0}  R_\Gamma^{(n)}\,
\Big(\frac{\alpha_s(\mu_i)}{4\pi}\Big)^n\,,
\nonumber \\
&~ \tilde{\phi}_+(\eta+a_\Gamma(\mu_i),\hat\omega,\mu_i)
& =&~
\sum_{n=0}  \tilde{\phi}_+^{(n)}\,
\Big(\frac{\alpha_s(\mu_i)}{4\pi}\Big)^n\,.
\end{align}
The scale $\mu_i$ is an initial scale, which should be $\mu_{h1}$, $\mu_{h2}$ and $\mu_{s}$ for evolution factors $U_1(E_\gamma,\mu_{h1},\mu)$, $U_2(\mu_{h2},\mu)$ and $\tilde{\phi}_{+,U}(\eta,\mu_s,\mu)$, respectively.

The expansion coefficients for $a_{\gamma_{\rm gen}}$ and $S$ are
\begin{eqnarray}
a_{\gamma_{\rm gen}}^{(0)}&=&~
\frac{{\gamma_{0,\rm gen}}}{2\beta_0}\,\ln r\,,
\quad
a_{\gamma_{\rm gen}}^{(1)}= \frac{1}{2\beta_0}\,
\Big({\gamma_{1,\rm gen}}
-\frac{{\gamma_{0,\rm gen}}\,\beta_1}{\beta_0}\Big)\,(r-1)\,,
\nonumber\\
a_{\gamma_{\rm gen}}^{(2)}&=&~
\frac{1}{4\beta_0}\,\Big({\gamma_{2,\rm gen}}
-\frac{{\gamma_{1,\rm gen}}\,\beta_1}{\beta_0}
+\frac{{\gamma_{0,\rm gen}}(\beta_1^2-\beta_0\,\beta_2)}
{\beta_0^2}\Big)\,(r^2-1)\,,
\\
&&\nonumber \\
S^{(0)}&=&~\frac{\Gamma_0}{4\beta_0^2}\,\Big[\frac{4\pi}{\alpha_s(\mu_i)}\,(1-\ln r-\frac{1}{r}) +\frac{\beta_1}{2\beta_0}\ln^2 r- \Big(\frac{\Gamma_1}{\Gamma_0} -\frac{\beta_1}{\beta_0}\Big)(r-1-\ln r)\Big]\,,
\nonumber\\
S^{(1)}&=&-\frac{\Gamma_0}{8\beta_0^2}\,\Big[ \frac{\Gamma_2}{\Gamma_0}(r-1)^2 +\frac{\beta_2}{\beta_0}(1-r^2+2\ln r)-\frac{\Gamma_1\beta_1}{\Gamma_0\beta_0}(3-4r+r^2+2r\ln r)
\nonumber\\
&+&\frac{\beta_1^2}{\beta_0^2}(1-r)(1-r-2\ln r) \Big]\,,
\nonumber\\
S^{(2)}&=&~\frac{\Gamma_0}{48\beta_0^3}\, \Big[ -\frac{2\Gamma_3\beta_0}{\Gamma_0}(r-1)^2(2r+1)
-\beta_3\,(1+3r^2-4r^3+6\ln r)
\nonumber\\
&+&\frac{\Gamma_2\beta_1}{\Gamma_0}\, (5-9r^2+4r^3+6r^2\ln r)
+\frac{\beta_2\beta_1}{\beta_0}\, (-7+12r+3r^2-8r^3+6(2-r^2)\ln r)
\nonumber\\
&+& \frac{4\Gamma_1\beta_2}{\Gamma_0}\, (2-3r+r^3)
+\frac{\Gamma_1\beta_1^2}{\Gamma_0\beta_0}\,(-11+12r+3r^2-4r^3-6r^2\ln r)
\nonumber\\
&+&\frac{2\beta_1^3}{\beta_0^2}\, (4-6r+2r^3+3(r^2-1)\ln r)\Big]\,,
\end{eqnarray}
with $r= \alpha_s(\mu)/\alpha_s(\mu_i)$.
Note that $S^{(0)}$ also includes an order $\alpha^{-1}_s$ term.
The expansion coefficients for $G$ are
\begin{eqnarray}
G^{(1)}&=&-\frac{\Gamma_0}{2\beta_0}\,r
\int_0^1dx\, \frac{x^{-\eta}}{1-x}\,h^{(1)}(x)\,
\frac{1-r^{-l_x-1}}{l_x+1}\,,
\nonumber\\
G^{(2)}&=&~\frac{\Gamma_0}{2\beta_0}\,r^2
\int_0^1dx\, \frac{x^{-\eta}}{1-x}\,
\bigg\{h^{(1)}(x)\, \bigg[
\bigg(\frac{\beta_1}{\beta_0}
- \Big( \frac{\Gamma_1}{\Gamma_0}
- \frac{\beta_1}{\beta_0} \Big)\,l_x\bigg)\,
\frac{1-r^{-l_x-2}}{l_x+2}
\nonumber\\
&-&\Big( \frac{\Gamma_1}{\Gamma_0}
- \frac{\beta_1}{\beta_0} \Big) \,\frac{l_x\,
\big(1-  r^{-l_x-1} \big)}{l_x+1} \bigg]-
h^{(2)}(x)\,\frac{1-r^{-l_x-2}}{l_x+2}\bigg\}\,,
\end{eqnarray}
where $l_x \equiv \Gamma_0/(2\beta_0) \ln x$.
The expansion coefficients for $R_\Gamma$ and $\tilde{\phi}_+$ are
\begin{align}
R^{(0)}_\Gamma =&~
\frac{\Gamma\big(1+\eta+a^{(0)}_\Gamma\big)\,\Gamma(1-\eta)}
{\Gamma\big(1-\eta-a^{(0)}_\Gamma\big)\,\Gamma(1+\eta)}\,,
\qquad
R^{(1)}_\Gamma =
a^{(1)}_\Gamma \,
\Big[\psi(1+\eta+a^{(0)}_\Gamma\big)
+ \psi(1-\eta-a^{(0)}_\Gamma\big)\Big] \,
R^{(0)}_\Gamma \,,
\nonumber \\
R^{(2)}_\Gamma =&~
\frac{a^{(2)}_\Gamma}{a^{(1)}_\Gamma}\,R^{(1)}_\Gamma
+ \frac{1}{2}\,\bigg\{
a^{(1)2}_\Gamma \, \Big[\psi^{(1)}(1+\eta+a^{(0)}_\Gamma\big)
- \psi^{(1)}(1-\eta-a^{(0)}_\Gamma\big)\Big] \,
+\Big(\frac{R^{(1)}_\Gamma}{R^{(0)}_\Gamma}\Big)^2
\bigg\} R^{(0)}_\Gamma\,,
\\
& \nonumber \\
\tilde{\phi}_+^{(0)} = &~ \tilde{\phi}_+(\eta+a^{(0)}_\Gamma)\,,
\quad
\tilde{\phi}_+^{(1)} =
a^{(1)}_\Gamma\, \tilde{\phi}'_+(\eta+a^{(0)}_\Gamma)\,,
\quad
\tilde{\phi}_+^{(2)} =~
a^{(2)}_\Gamma\, \tilde{\phi}'_+(\eta+a^{(0)}_\Gamma)
+\frac{a^{(1)2}_\Gamma}{2}\,
\tilde{\phi}''_+(\eta+a^{(0)}_\Gamma) \,.
\end{align}
The expansion coefficients of $a_\Gamma\equiv a_{\Gamma_{\rm cusp}}$ can be obtained from that of $a_{\gamma_{\rm gen}}$ by a replacement of the anomalous dimension $\gamma_{\rm gen}\to \Gamma_{\rm cusp}$.

\section{Photon LCDA}\label{DAofPhoton}
We summarize the operator-level definitions of the two-particle and three-particle photon distribution amplitudes
up to and including the twist-four accuracy, following the systematic classification detailed in~\cite{Ball:2002ps},
and taking this opportunity to correct several misprints in the previous expressions
displayed in Appendix B of~\cite{Wang:2018wfj}.
\begin{eqnarray}
&& \langle \gamma(p) |\bar q(x) \, W_c(x, 0) \,\, \sigma_{\alpha \beta} \,\, q(0)| 0 \rangle  \nonumber \\
&& = - i \, g_{\rm em} \, Q_q \,  \langle \bar q q \rangle(\nu) \,
(p_{\beta} \, \epsilon_{\alpha}^{\ast} - p_{\alpha} \, \epsilon_{\beta}^{\ast}) \,
\int_0^1 \, d u \, e^{i \, u \, p \cdot x} \, \left [  \chi(\nu) \, \phi_{\gamma}(u, \nu)
+ {x^2 \over 16} \, \mathbb{A}(u, \nu) \right ]   \nonumber \\
&& \hspace{0.3 cm} - \, {i \over 2} \, g_{\rm em} \, Q_q \,  {\langle \bar q q \rangle(\nu) \over q \cdot x} \,
(x_{\beta} \, \epsilon_{\alpha}^{\ast} - x_{\alpha} \, \epsilon_{\beta}^{\ast}) \,
\int_0^1 \, d u \, e^{i \, u \, p \cdot x} \, h_{\gamma}(u, \nu) \,, \\
\nonumber \\
&& \langle \gamma(p) |\bar q(x) \, W_c(x, 0) \,\, \gamma_{\alpha} \,\, q(0)| 0 \rangle
=  g_{\rm em} \, Q_q \, f_{3 \gamma}(\nu) \, \epsilon_{\alpha}^{\ast} \,
\int_0^1 \, d u \, e^{i \, u \, p \cdot x} \, \psi^{(v)}(u, \nu)  \,, \\
\nonumber \\
&& \langle \gamma(p) |\bar q(x) \, W_c(x, 0) \,\, \gamma_{\alpha} \, \gamma_5 \,\, q(0)| 0 \rangle \nonumber \\
&& = {g_{\rm em} \over 4} \, Q_q \,  f_{3 \gamma}(\nu) \, \varepsilon_{\alpha \beta \rho \tau} \,
p^{\rho} \, x^{\tau} \,\epsilon^{\ast \, \beta} \, \int_0^1 \, d u \, e^{i \, u \, p \cdot x} \,
\, \psi^{(a)}(u, \nu)\,, \\
\nonumber \\
&& \langle \gamma(p) |\bar q(x) \, W_c(x, 0) \,\, g_s \, G_{\alpha \beta}(v \, x) \, \, q(0)| 0 \rangle \nonumber \\
&& =  - i \, g_{\rm em} \, Q_q \,  \langle \bar q q \rangle(\nu) \,
(p_{\beta} \, \epsilon_{\alpha}^{\ast} - p_{\alpha} \, \epsilon_{\beta}^{\ast}) \,
\int [{\cal D} \underline{\alpha}] \, e^{i \, (\alpha_q +  \, v \, \alpha_g) \, p \cdot x} \, S(\underline{\alpha}; \nu) \,, \\
\nonumber \\
&& \langle \gamma(p) |\bar q(x) \, W_c(x, 0) \,\, g_s \, \widetilde{G}_{\alpha \beta}(v \, x) \,
i \, \gamma_5 \,\,  q(0)| 0 \rangle \nonumber \\
&& =  i \, g_{\rm em} \, Q_q \,  \langle \bar q q \rangle(\nu) \,
(p_{\beta} \, \epsilon_{\alpha}^{\ast} - p_{\alpha} \, \epsilon_{\beta}^{\ast}) \,
\int [{\cal D} \underline{\alpha}] \, e^{i \, (\alpha_q +  \, v \, \alpha_g) \, p \cdot x} \, \widetilde{S}(\underline{\alpha}; \nu) \,, \\
\nonumber \\
&& \langle \gamma(p) |\bar q(x) \, W_c(x, 0) \,\, g_s \, \widetilde{G}_{\alpha \beta}(v \, x) \,
\gamma_{\rho} \, \gamma_5 \,\,  q(0)| 0 \rangle \nonumber \\
&& = - g_{\rm em} \, Q_q \, f_{3 \gamma}(\nu) \, p_{\rho} \,
(p_{\beta} \, \epsilon_{\alpha}^{\ast} - p_{\alpha} \, \epsilon_{\beta}^{\ast}) \,
\int [{\cal D} \underline{\alpha}] \, e^{i \, (\alpha_q +  \, v \, \alpha_g) \, p \cdot x} \, A(\underline{\alpha}; \nu) \,, \\
\nonumber \\
&& \langle \gamma(p) |\bar q(x) \, W_c(x, 0) \,\, g_s \, G_{\alpha \beta}(v \, x) \,
i \, \gamma_{\rho} \,\,  q(0)| 0 \rangle \nonumber \\
&& = - g_{\rm em} \, Q_q \, f_{3 \gamma}(\nu) \,  p_{\rho} \,
(p_{\beta} \, \epsilon_{\alpha}^{\ast} - p_{\alpha} \, \epsilon_{\beta}^{\ast}) \,
\int [{\cal D} \underline{\alpha}] \, e^{i \, (\alpha_q +  \, v \, \alpha_g) \, p \cdot x} \, V(\underline{\alpha}; \nu)  \,, \\
\nonumber \\
&& \langle \gamma(p) |\bar q(x) \, W_c(x, 0) \,\, g_{\rm em} \, Q_q \,  F_{\alpha \beta}(v \, x) \, \, q(0)| 0 \rangle \nonumber \\
&& =  - i \, g_{\rm em} \, Q_q \,  \langle \bar q q \rangle(\nu) \,
(p_{\beta} \, \epsilon_{\alpha}^{\ast} - p_{\alpha} \, \epsilon_{\beta}^{\ast}) \,
\int [{\cal D} \underline{\alpha}] \, e^{i \, (\alpha_q +  \, v \, \alpha_g) \, p \cdot x} \, S_{\gamma}(\underline{\alpha}; \nu) \,,  \\
\nonumber \\
&& \langle \gamma(p) |\bar q(x) \, W_c(x, 0) \,\, \sigma_{\rho \tau} \,\, g_s \, G_{\alpha \beta}(v \, x)
\,\,  q(0)| 0 \rangle \nonumber \\
&& =  \, g_{\rm em} \, Q_q \,\langle \bar q q \rangle(\nu) \,
\left [p_{\rho} \, \epsilon_{\alpha}^{\ast} \, g_{\tau \beta}^{\perp}
- p_{\tau} \, \epsilon_{\alpha}^{\ast} \, g_{\rho \beta}^{\perp} - (\alpha \leftrightarrow \beta) \right ]  \,
\int [{\cal D} \underline{\alpha}] \, e^{i \, (\alpha_q +  \, v \, \alpha_g) \, p \cdot x} \, T_{1}(\underline{\alpha}; \nu)  \nonumber \\
&& \hspace{0.4 cm} + \, g_{\rm em} \, Q_q \,\langle \bar q q \rangle(\nu) \,
\left [p_{\alpha} \, \epsilon_{\rho}^{\ast} \, g_{\tau \beta}^{\perp}
- p_{\beta} \, \epsilon_{\rho}^{\ast} \, g_{\tau \alpha}^{\perp} - (\rho \leftrightarrow \tau) \right ]  \,
\int [{\cal D} \underline{\alpha}] \, e^{i \, (\alpha_q +  \, v \, \alpha_g) \, p \cdot x} \, T_{2}(\underline{\alpha}; \nu) \nonumber \\
&& \hspace{0.4 cm} + \, g_{\rm em} \, Q_q \,\langle \bar q q \rangle(\nu) \,
\frac{(p_{\alpha} \, x_{\beta} - p_{\beta} \, x_{\alpha} ) (p_{\rho} \, \epsilon_{\tau}^{\ast} - p_{\tau} \,\epsilon_{\rho}^{\ast})}
{p \cdot x} \, \int [{\cal D} \underline{\alpha}] \, e^{i \, (\alpha_q +  \, v \, \alpha_g) \, p \cdot x} \,
 T_{3}(\underline{\alpha}; \nu) \nonumber \\
&& \hspace{0.4 cm}  + \, g_{\rm em} \, Q_q \,\langle \bar q q \rangle(\nu) \,
\frac{(p_{\rho} \, x_{\tau} - p_{\tau} \, x_{\rho} ) (p_{\alpha} \, \epsilon_{\beta}^{\ast} - p_{\beta} \,\epsilon_{\alpha}^{\ast})}
{p \cdot x} \, \int [{\cal D} \underline{\alpha}] \, e^{i \, (\alpha_q +  \, v \, \alpha_g) \, p \cdot x} \,
 T_{4}(\underline{\alpha}; \nu) \,, \\
 \nonumber \\
&& \langle \gamma(p) |\bar q(x) \, W_c(x, 0) \,\, \sigma_{\rho \tau} \,\, g_{\rm em} \, Q_q \, F_{\alpha \beta}(v \, x)
\,\,  q(0)| 0 \rangle \nonumber \\
&& =  \, g_{\rm em} \, Q_q \,\langle \bar q q \rangle(\nu) \,
\frac{(p_{\rho} \, x_{\tau} - p_{\tau} \, x_{\rho} ) (p_{\alpha} \, \epsilon_{\beta}^{\ast} - p_{\beta} \,\epsilon_{\alpha}^{\ast})}
{p \cdot x} \, \int [{\cal D} \underline{\alpha}] \, e^{i \, (\alpha_q +  \, v \, \alpha_g) \, p \cdot x} \,
 T_{4}^{\gamma}(\underline{\alpha}; \nu)  + \cdots .
\end{eqnarray}
Here, we adopt the following convention for the dual gluon-field strength tensor
\begin{eqnarray}
\widetilde{G}_{\alpha \beta}= {1 \over 2} \,  \varepsilon_{\alpha \beta \rho \tau }  \, G^{\rho \tau} \,,
\qquad
\int [\mathcal{D}\underline{\alpha}]
= \int d\alpha_q \int d\alpha_{\bar{q}} \int d\alpha_g\,
\delta(1-\alpha_q-\alpha_{\bar{q}}-\alpha_g) \,,
\end{eqnarray}
with $\underline{\alpha}=\{\alpha_q,\alpha_{\bar{q}},\alpha_g\}$, and $\alpha_q,~\alpha_{\bar{q}},~\alpha_g$ are momentum fractions carried by the quark, anti-quark and gluon in the photon, respectively.
It is convenient to define $\bar{\psi}$ and $\bar{h}_\gamma$  by
\begin{eqnarray}
\bar{\psi}^{(v)}(u,\nu)=2\int_0^ud\hat{u}\,
\psi^{(v)}(\hat{u},\nu)\,, \quad
\bar{h}_{\gamma}(u,\nu)=-4\int_0^ud\hat{u}
\,(u-\hat{u})\,h_\gamma(\hat{u},\nu)\,.
\end{eqnarray}

\bibliography{references}
\bibliographystyle{apsrev4-1}
\end{document}